\documentclass[aip, reprint]{revtex4-2}

\usepackage{XiGrpCommon}
\usepackage{subcaption}
\usepackage{amsmath}
\usepackage{graphicx}
\usepackage{comment}
\usepackage{mathtools}
\usepackage{siunitx}
\graphicspath{{./figures/}} 

\usepackage{CJKutf8}

\renewcommand{\RevisedText}[1]{#1}

\begin{document}
	\begin{CJK*}{UTF8}{gbsn}
		\CJKtilde 
		\CJKindent 
		
		\title{Equilibrium and non-equilibrium molecular dynamics approaches for the linear viscoelasticity of polymer melts}
		\author{Oluseye Adeyemi}
		\affiliation{Department of Chemical Engineering, McMaster University, Hamilton, Ontario L8S 4L7, Canada}
		\author{Shiping Zhu (朱世平)}
		\affiliation{Department of Chemical Engineering, McMaster University, Hamilton, Ontario L8S 4L7, Canada}
		\author{Li Xi (奚力)}
		\email[coresponding author, E-mail: ]{xili@mcmaster.ca}
		\homepage[Web: ]{https://www.xiresearch.org}
		\affiliation{Department of Chemical Engineering, McMaster University, Hamilton, Ontario L8S 4L7, Canada}
		\affiliation{School of Computational Science and Engineering, McMaster University, Hamilton, Ontario L8S 4K1, Canada}
		\newpage
\begin{abstract}
Viscoelastic properties of polymer melts are particularly challenging to compute due to the intrinsic stress fluctuations in molecular dynamics (MD). We compared equilibrium and non-equilibrium MD approaches for extracting the storage ($G'$) and loss moduli ($G''$) over a wide frequency range from a bead-spring chain model, in both unentangled and entangled regimes. We found that, with properly chosen data processing and noise reduction procedures, different methods render quantitatively equivalent results. In equilibrium MD (EMD), applying the Green-Kubo relation with a multi-tau correlator method for noise filtering generates smooth stress relaxation modulus profiles, from which accurate $G'$ and $G''$ can be obtained. For unentangled chains, combining the Rouse model with a short-time correction provides a convenient option that circumvents the stress fluctuation challenge altogether. For non-equilibrium MD (NEMD), we found that combining a stress pre-averaging treatment with discrete Fourier transform analysis reliably computes $G'$ and $G''$ with much shorter simulation length than previously reported. Comparing the efficiency and statistical accuracy of these methods, we concluded that EMD is both reliable and efficient, and is suitable when the whole spectrum of linear viscoelastic properties is desired, whereas NEMD offers flexibility when only some frequency ranges are of interest.
\end{abstract}
\maketitle
\end{CJK*}

\newpage
\section{Introduction}\label{Sec_intro}
The linear viscoelastic (LVE) properties of polymers provide unique insights into their structure and also govern the flow behavior during processing. These properties are usually measured by a small displacement of the polymer molecules from their equilibrium positions, thereby ensuring that the response is still in the linear regime. Experimentally, LVE properties are determined by a small amplitude oscillatory shear (SAOS) experiment\cite{ferry1980viscoelastic,bird1987dynamics}, which provides the storage ($G'$) and loss ($G''$) moduli of the material over a frequency spectrum. The accessible frequency ranges are either limited by the equipment capabilities or the degradation of the polymers at high shear rates or temperatures. The high-shear rate challenge is typically mitigated by the temperature superposition technique. \par
Owing to the range of time and length scales involved, computing the viscoelastic properties of polymers in molecular dynamics (MD) simulations still remains a formidable task. Indeed, for long-chain polymers, it remains unrealistic to capture the whole spectrum of linear viscoelasticity using fully atomistic molecular models. Even for highly coarse-grained models, accurate determination of viscoelastic properties in MD must still overcome the challenges of long relaxation times and strong stress fluctuations. In particular, for highly-entangled polymers, MD must be combined with high-level polymer dynamics models for quantitative prediction~\cite{xi2019molecular}.\par 
Regardless of the model being used, extraction of viscoelastic properties from MD simulations is an essential step. This can be achieved with either equilibrium or non-equilibrium MD (EMD and NEMD) simulation. The EMD approach samples the spontaneous stress fluctuations in the thermodynamic ensemble of the system. The shear stress relaxation modulus $G(t)$, from which linear viscoelastic material functions are calculated, is related to the time autocorrelation function (TACF) of the stress tensor through the Green-Kubo (GK) relation~\cite{j2007statistical}. The NEMD approach, on the other hand, models the flow condition of rheological measurement from which the corresponding material function is directly computed~\cite{cummings1992nonequilibrium}.\par
Take shear viscosity, which is the most computed rheological property in the literature, for example. Since the EMD approach simulates equilibrium conditions, it can only provide the zero-shear viscosity as a temporal integral of the relaxation modulus
\begin{gather}
	\eta_0\equiv\lim_{\dot\gamma\to 0}\eta=\int_0^\infty G(t)dt
	\label{eq:eta0}
\end{gather}
(where $\dot\gamma$ is the shear rate). By constrast, the NEMD approach simulates the steady shear flow condition and calculates the viscosity by dividing the steady-state shear stress by the shear rate. For simple liquids such as the Lennard-Jones (LJ) fluid, shear viscosity values from EMD and NEMD approaches agree well~\cite{hess2002determining, chen2009pressure}. Although there was a general perception that the EMD approach is prone to large statistical uncertainty due to intense stress fluctuations in molecular systems and difficult convergence of the integral in \cref{eq:eta0}, it has been shown that reliable results are attainable with careful selection of the integration limits and data processing procedure \cite{chen2009pressure, zhang2015reliable}. For polymers, viscosity is in general a function of shear rate, but a Newtonian plateau exists at the small $\dot\gamma$ limit. Extrapolation of the $\eta(\dot\gamma)$ profile from NEMD to the $\dot\gamma=0$ limit again agrees well with the EMD value from \cref{eq:eta0}~\cite{kroger2000rheological, xu1995transport, xu1997calculation, sen2005viscoelastic, vladkov2006linear}. \par 
The focus of this study is on the full spectrum of linear viscoelastic properties as reflected in the frequency ($\omega$)-dependent $G'(\omega)$ and $G''(\omega)$ profiles. Compared with shear viscosity, the computational cost for obtaining $G'$ and $G''$ is significantly higher (in both EMD and NEMD) as viscoelastic responses at a wide range of frequencies are now required. The EMD approach again relies on the GK relation and was first reported by \citet{sen2005viscoelastic}, followed by a number of later developments \cite{likhtman2007linear, lee2009entangled}. Many more studies reported the EMD results of $G(t)$ but did not convert it to $G'(\omega)$ and $G''(\omega)$ \cite{zhou2006direct, hou2010stress, hsu2016static, peters2020viscoelastic, adeyemi2021dynamics}. The NEMD approach simulates the sinusoidal oscillatory shear flow (modeling the SAOS condition) and obtains $G'$ and $G''$ from the time-dependent shear stress signal. Those efforts date back to earlier studies by \citet{cifre2004linear} and \citet{vladkov2006linear}, and NEMD results of $G'$ and $G''$ were also reported more recently by \citet{karim2016determination}. \par
Information on the comparsion between these two approaches is rather limited. For the bead-spring chain or Kremer-Grest (KG) model \cite{kremer1990dynamics} of very short chain length ($N=10$ and $20$), \citet{vladkov2006linear} conducted a comparative study between EMD and NEMD approaches for polymer viscoelasticity. Direct comparison between the GK and NEMD approaches, however, was only reported for the zero-shear viscosity. The study did not report the $G'$ and $G''$ results from the GK relation because it was not able to extract statistically meaningful results buried under strong noises.
It instead proposed a corrected Rouse mode analysis (cRMA) approach which brings in MD data to fill in the short-time dynamics missing in the Rouse model. The method is fundamentally still an EMD approach but its viability relies on the accuracy of the Rouse model which is designed only for unentangled polymers. For the short chains studied, good agreement was found between $G'$ and $G''$ results from cRMA and NEMD.
More recently, Karim et al.\cite{karim2012determination, karim2016determination} compared NEMD results of $G'$ and $G''$, for $N=20$ and $80$, from several sources with the GK results of \citet{sen2005viscoelastic}. Good agreement is generally found in the frequency range tested by NEMD (typically fewer than three decades) with discrepancy sometimes observed at the low frequency (long-time) limit where statistical uncertainty is highest in both methods. \par
The purpose of this study is to determine which method is better for the accurate calculation of $G'$ and $G''$ over a wide frequency range. Many researchers seem to prefer NEMD because of the general belief that EMD is more affected by strong stress fluctuations. Indeed, the large noise-to-signal ratio in the long-time tail of $G(t)$ obtained from the GK relation has sometimes caused erroneous conclusions in previous studies\cite{xi2019molecular}. It takes extremely long EMD simulations to effectively reduce the statistical uncertainty in $G(t)$. For shear viscosity calculation, it is widely accepted that NEMD requires substantially less computational cost for satisfactory accuracy \cite{hess2002determining, chen2009pressure, mondello1997viscosity, xi2019molecular}. We note that this advantage does not straightforwardly translate to $G'(\omega)$ and $G''(\omega)$ calculation because NEMD must be separately performed for each frequency level of interest, while the EMD approach allows the calculation of the whole spectrum with one long simulation run. \par
In this study, we directly compare the accuracy and efficiency of EMD and NEMD approaches for computing $G'(\omega)$ and $G''(\omega)$ profiles.
This is the first time these two approaches are compared with identical molecular models, which will allow us to identify the discrepancies, if any, that are attributed solely to the difference in the methodology for computing viscoelastic properties.
In addition to evaluating the quantitative equivalence between their results, efficiency, in terms of which method provides statistically more accurate results with limited computational resources, is also a key consideration. For EMD, our primary focus is on the GK approach, but we also include the cRMA approach for completeness. Methods are evaluated in monodisperse melts of KG chains with $N=25$ to $350$, covering both unentangled and (moderately) entangled regimes. To our knowledge, NEMD calculation of $G'$ and $G''$ has not been previously reported for entangled polymers. Both categories of methods are strongly influenced by statistical errors due to stress fluctuations in MD simulation.
We have experimented with various noise reduction techniques and present the best procedure that we find for each method. This allows us to compare the methods on an equal footing -- i.e., each method is evaluated at its optimal settings. Therefore, in addition to guiding the choice of method for computing viscoelastic properties, the study also aims to demonstrate the best practice in each approach.  

\section{Methods}\label{sec:methods}
\subsection{Simulation Details}
We model the polymer chains using the classical Kremer-Grest (KG) bead-spring chain model\cite{kremer1990dynamics}. Consecutive beads in a polymer chain interact with the finitely extensible non-linear elastic (FENE) springs potential
\begin{align}
	U_\text{FENE}(r) &= -\frac{1}{2}K{R_0}^2\text{ln}\left[1-{\left(\frac{r}{R_0}\right)}^2\right] \notag \\
	&+4\epsilon\left[{\left(\frac{\sigma}{r}\right)}^{12}-{\left(\frac{\sigma}{r}\right)}^6+\frac{1}{4}\right]
\end{align}
where $r$ represents the distance between the beads, and $\sigma$ and $\epsilon$ are the LJ length and energy parameters. The first term of the equation models an attractive potential due to the entropic interaction between the polymer segments, which diverges at a maximum bond length $R_0 = 1.5\sigma$. The second term represents the repulsive force between beads and is only included at distances $r \leq 2^{\frac{1}{6}} \sigma$. The spring force $K = 30 \epsilon/\sigma^2$ allows the use of a large integration timestep and also prevents the bonds from cutting through each other. The interaction potential between the non-bonded beads is modeled by the standard Lennard Jones (LJ) potential
\begin{equation}\label{eq:LJ}
	U_\mathrm{LJ}(r)=4\epsilon \left(\left(\frac{\sigma}{r}\right)^{12} - \left(\frac{\sigma}{r}\right)^{6} \right) 
\end{equation}
for which a cutoff of $2.5\sigma$ is used and a vertical offset is added to ensure continuity at the cutoff. All the results are reported in reduced LJ units and length, energy, time, and temperature values are scaled by, $\sigma$, $\epsilon$, $\tau = \sqrt{m\sigma^2/\epsilon}$, and $\epsilon/k_B$ ($k_B$ is the Boltzmann constant) respectively.

The chain lengths studied range from the unentangled $N=25$ and $50$ to marginally entangled $N=100$ and moderately entangled $N=350$ cases\cite{adeyemi2021dynamics}. The $N=350$ case contains a total of $56000$ beads in the simulation box while all other cases contain $50000$ beads in each simulation box. All simulations were performed at a constant bead density of 0.85 $\sigma^{-3}$.
\RevisedText{%
The corresponding simulation box size, measured by the length of each edge, ranges from $38.90\sigma$ (for \num{50000} beads) to $40.38\sigma$ (for \num{56000} beads).
In \citet{adeyemi2021dynamics}, we have reported the Flory's characteristic ratio $C_\infty = 1.75$ for the KG polymer melt used in our study.
The mean end-to-end distance $R$ of the chains can be calculated using $R^2 = C_\infty nr_\text{b}^2$, where $n=N-1$ is the number of bonds and $r_\text{b}=0.97\sigma$ is the equilibrium FENE bond length.
Even for $N=350$ -- i.e., the longest chains studied, the estimated $R=23.90\sigma$, which is still sufficiently short, in comparison with the box dimension, to prevent interaction between periodic images of the same chain.
}%
The temperature of the simulations was maintained at $1\epsilon/k_B$ with Nos{\'e}-Hoover chains.

All the simulations were carried out using the Large-scale Atomic/Molecular Massively Parallel Simulator (LAMMPS) package\cite{plimpton1995fast}. The equation of motion was integrated using the velocity Verlet algorithm with a time step of $\Delta t = 0.01$ (in LJ time units or TUs).
\RevisedText{For selected cases, we have repeated the simulation with $\Delta t=0.005$ and confirmed that the results do not depend on the time step size.}

Initial configurations were generated by randomly placing the specified number and types of chains in a cell following a self-avoiding walk conformation statistics. The structures were further equilibrated using a modified dissipative particle dynamics (DPD) push-off step \citep{sliozberg2012fast} during which a soft repulsive potential 
\begin{align}
	U_\text{DPD}(r) = 
	\begin{dcases}
		\frac{A_\text{DPD}}{2}r_c(1-\frac{r}{r_c}), & (r < r_c) \\
		0, & (r \ge r_c)
	\end{dcases}
\end{align}
was used to replace the LJ potential (\cref{eq:LJ}) between the non-bonded beads. DPD equilibration was performed at $T=1.0$ and used a cut-off distance $r_c = 1.0$. The DPD potential was initially kept low at $A_\text{DPD} = 25$. At the beginning, we restricted the maximum distance that each bead can move in a single time step and gradually increased it from $0.001$ to $0.1$ over $15\mathrm{TUs}$. After the restriction was removed, we further ran the DPD simulation for another $100\mathrm{TUs}$, following which  $A_\mathrm{DPD}$ was gradually ramped up to $100$ over $5\mathrm{TUs}$. Finally, we replaced the DPD potential with the standard LJ potential (\cref{eq:LJ}) and performed MD simulation in an NVT ensemble for another $500\mathrm{TUs}$ during which a random velocity distribution was assigned to all the beads every $0.5\mathrm{TUs}$. Mean square internal displacement of the chains, which is a sensitive indicator of unrelaxed chain conformations\cite{auhl2003equilibration}, was examined to ensure the convergence of the equilibration procedure -- see \citet{adeyemi2021dynamics}.
\subsection{Equilibrium Molecular Dynamics (EMD) or Green-Kubo (GK) Approach}\label{sec:EMD}
The GK relation relates the shear stress relaxation modulus $G(t)$ to the TACF of shear stress fluctuations
\begin{equation}\label{eq:GK}
	G(t) = \frac{V}{k_B T}\langle \sigma_{xy}(t)\sigma_{xy}(0) \rangle
\end{equation} 
where $V$ is the volume of the system, $T$ is the temperature and $\sigma_{xy}$ is an off-diagonal stress component. The major challenge in using this approach is the intense fluctuations of the stress TACF which is particularly severe at the terminal (large $t$) regime. One strategy for the reduction of fluctuation is by pre-filtering the stress signal with moving average before the TACF is calculated\cite{masoori2017reducing,hsu2016static}. Alternatively, moving average may be applied directly to the $G(t)$ profile\cite{sen2005viscoelastic}. The window size for moving average must be carefully selected to prevent the data from being overly smeared. \citet{lee2009entangled} found that $G(t)$ calculated from the filtered $\sigma_{xy}(t)$ signal is artificially reduced at the short-time end, but argued that, with properly-chosen window size, the long-time behavior of $G(t)$ is unaffected.
Nevertheless, using a fixed window size in the moving average approach is intrinsically limited because not only are the fluctuations coming from various frequencies, but the uncertainty in $G(t)$ also grows with the time lag $t$ due to the diminishing number of independent segments for averaging in a fixed-length time series. For this reason, strong fluctuations at the long-time limit of $G(t)$ cannot be effectively tamed with moving average\cite{lee2009entangled, padding2002time, masoori2017reducing} which is often a cause of erroneous results\cite{xi2019molecular}.\par
A more delicate multi-tau correlator method, proposed by \citet{ramirez2010efficient}, was used in this study. From our practical experience, the method generates adequately smooth $G(t)$ profile across nearly the whole range of time lag except at the very long time end where the relaxation modulus has nearly vanished. The idea is to filter the stress signal $\sigma_{xy}(t)$ and calculate its TACF on the fly with a multi-level hierarchical data structure. Each level contains $p$ data points.
Level 0 stores the most recent $p$ points from the time series, from which TACF for time at $t = 0\Delta t, 1\Delta t \cdots, (p-1)\Delta t$, is calculated and also stored. At level $l$ ($l \geq 1$), each data entry is the average between $m$ data points from level $l-1$ and the most recent $p$ block averages (each covers $m^l$ data points in the original time series) are stored. Correspondingly, the TACF stored at each level also covers longer time lag than the previous one. Effectively, this method filters $\sigma_{xy}(t)$ with progressively larger window size for the TACF calculation at longer time lags. We used $m=2$ and $p=16$ as recommended by \citet{ramirez2010efficient}. \par
The equivalence between shear stress components of different directions in an isotropic fluid is also leveraged to reduce statistical error. Average over TACFs of those equivalent components is expected to have lower uncertainty than that of a single component $\sigma_{xy}$\cite{daivis1994comparison}. The particular form used in this study 
\begin{align}
	G(t) &{}=  \frac{V}{5k_BT}[\langle\sigma_{xy}(t)\sigma_{xy}(0)\rangle+\langle\sigma_{yz}(t)\sigma_{yz}(0)\rangle \notag \\
	&\qquad+\langle\sigma_{zx}(t)\sigma_{zx}(0)\rangle] \notag \\ &{}+\frac{V}{30k_BT}[\langle N_{xy}(t)N_{xy}(0)\rangle +\langle N_{xz}(t)N_{xz}(0)\rangle \notag \\ 
	&\qquad+ \langle N_{yz}(t)N_{yz}(0)\rangle]
\end{align}
where
\begin{gather}
	N_{\alpha \beta}=\sigma_{\alpha \alpha}-\sigma_{\beta \beta}
\end{gather}
($\alpha,\beta=x,y,z$) is the same as that used in \citet{ramirez2010efficient}.\par
Combining these measures allowed us to produce an adequately smooth $G(t)$ for the computation of the dynamic moduli $G'$ and $G''$ through 
\begin{equation} \label{eq:Gprime}
	G' = \omega \int_{0}^{\infty}G(t)\sin(\omega t) dt\\
\end{equation}
and
\begin{equation} \label{eq:Gdprime}
	G'' =  \omega \int_{0}^{\infty}G(t)\cos(\omega t) dt.
\end{equation}
Numerical evaluation of \cref{eq:Gprime} and \cref{eq:Gdprime} is not as straightforward as it may appear, because the multi-tau correlator method returns $G(t)$ on a non-uniform grid: the spacing between consecutive points increases with time lag $t$. \citet{likhtman2007linear} fitted the $G(t)$ profile to a series of Maxwell modes, from which the integrals were evaluated analytically.
The Maxwell modes approximate $G(t)$ with the superposition of exponential decay functions, which thus cannot capture oscillations in the profile.
We used a different approach and approximated $G(t)$ with piecewise linear functions and integrated each piece analytically. With sufficient resolution, this treatment retains all the variations in the $G(t)$ profile while also avoiding nonlinear regression. Details of our method are given in Appendix \ref{appendix_A}.\par
As listed in \cref{tab:timescales}, multiple separate EMD simulation runs were performed for each case and the average of those independent runs was reported. The duration of each independent simulation run matches that of the corresponding chain length in \citet{likhtman2007linear}
\begin{table}
\centering
\caption{EMD simulation parameters, including the duration of each independent simulation and number of independent simulations used. 
The maximum stress relaxation time $\tau_\text{max}$ is defined as the time when the obtained $G(t)$ (\cref{fig:Gt_plot}) decays to $10^{-3}$.%
}
\label{tab:timescales}
\begin{tabular}{cccc}
	\hline
	\hline
	$N$	&Simulation Duration (TUs) &Num. Runs
	&$\tau_\text{max}$ (TUs)	\\
	\hline
	25		&$\num{5e5}$	&5	&$\num{1.065e3}$	\\
	50		&$\num{5e5}$	&5	&$\num{2.949e3}$	\\
	100		&$\num{1e6}$	&5	&$\num{1.835e4}$	\\
	350		&$\num{3e6}$	&3	&$\num{4.614e5}$	\\
	\hline
	\hline
\end{tabular}
\end{table}

The EMD approach is particularly appealing because one simulation run contains the information for the whole LVE profile.
Meanwhile, if information is desired outside the linear regime, NEMD would be the only viable approach.

\subsection{Non-Equilibrium Molecular Dynamics (NEMD) Approach}\label{sec:NEMD}
The NEMD technique measures the system's unsteady response to an induced perturbation. Unlike the EMD method, this approach mimics a real experimental setup by imposing the corresponding flow condition on the simulation box. In the determination of $G'$ and $G''$, the deformation is SAOS.
The SLLOD equations of motion were used, which imposes a time-dependent velocity profile across the domain\cite{evans1984nonlinear}. The imposed velocity corresponds to a sinusoidal strain of 
\begin{equation}\label{eq:strain}
	\gamma(\omega) = \gamma_0 \sin (\omega t)
\end{equation}
where $\gamma_0$ is the amplitude of the oscillation and $\omega$ is the angular frequency. At the start of the simulation, an initial mean velocity profile that matches the instantaneous box deformation rate of the moment is imposed on all beads for the quick convergence of the flow condition. In general, for a viscoelastic sample, the stress response $\sigma (t)$ oscillates with the same frequency as the strain input
\begin{equation}
	\sigma(t)= \sigma_0 \sin(\omega t + \delta) \label{eq:sig1}.\\	
\end{equation}
There is, however, a phase angle shift $\delta$ which varies between $0$ and $\pi/2$ (purely elastic and purely viscous limits, respectively). The stress can be further decomposed into two orthogonal functions
\begin{equation}
	\sigma(t) = \gamma_0[G'(\omega)\sin(\omega t)+G''(\omega) \cos(\omega t)] \label{eq:sig2}
\end{equation}
such that one of them is in sync with the imposed strain (\cref{eq:strain}) and the other has a $\pi/2$ phase lead. \Cref{eq:sig2} above is easily seen from the trigonometric expansion of \cref{eq:sig1} using
\begin{equation}\label{eq:trig_exp}
	\sin (\omega t + \delta) = \cos \delta \sin (\omega t) + \sin \delta \cos(\omega t).
\end{equation}
Comparing \cref{eq:sig1}, \cref{eq:sig2}, and \cref{eq:trig_exp}, we get
\begin{align}
	&G' = \frac{\sigma_0}{\gamma_0} \cos \delta\\
	&G'' = \frac{\sigma_0}{\gamma_0} \sin \delta.
\end{align}
\par

Data processing for the NEMD method can also present significant challenges as the obtained $\sigma(t)$ time series is again loaded with strong noises. Previous studies often used least-square fitting of the NEMD stress output to obtain $G'$ and $G''$ in \cref{eq:sig2}\cite{cifre2004linear, tseng2010linear}. In Appendix \ref{appendix_B}, we show that, in the absence of noise, a simple discrete Fourier transform (DFT) of the sinusoidal time series
\begin{gather}
	s(t)\equiv\frac{\sigma(t)}{\gamma_0}
\end{gather}
only has two non-zero modes
\begin{gather}\label{eq:gg_fourier}
	\hat{s}_{\pm k_\omega} = \frac{1}{2}\left(G''\mp iG'\right)
\end{gather}
where $\hat\cdot$ denotes Fourier modes and $k_\omega$ is the wavenumber corresponding to the imposed frequency $\omega$: i.e.,
\begin{gather}
	k_\omega = N_\text{cycle}
\end{gather}
is the number of complete oscillatory cycles in the simulation run. Stress fluctuations from simulation will show up in a wide range of frequencies, but the signal at the $\pm k_\omega$ modes will still be the dominant ones and their imaginary and real parts are related to $G'$ and $G''$, respectively. In practice, we additionally performed a noise-filtering step by pre-averaging the $\sigma(t)$ signal before the DFT analysis. The $\sigma(t)$ time series was divided into small blocks, each of which covers $1/100$ of an oscillatory cycle. The average of each block was used to compute $s(t)$ -- the input of DFT.
Since the block size and oscillatory cycle differ by two orders of magnitude, this step is designed to smoothen the signal without interfering with the primary Fourier modes. Applying DFT directly on the NEMD stress output without pre-averaging, according to our tests, will give nearly identical $G'$ and $G''$ at high frequencies. At low frequencies, however, its results contain strong, seemingly random, statistical errors. \par
We performed NEMD for 50 frequency levels spanning four decades of $\omega$ (from $10^{-4}$ to $1$). Simulation at each frequency level contains $N_\text{cycle}=25$ complete cycles. In total, $9.16\times10^{8}$ MD time steps were used for the entire spectrum. The number of time steps spent at each frequency level increases $\propto 1/\omega$. For comparison, $N_\text{cycle}=100$ to $200$ was often used in previous studies\cite{cifre2004linear, tseng2010linear}. As we will show in this study, with the noise reduction procedure described above, $N_\text{cycle}=25$ was sufficient to generate statistically robust results. Shortening of individual NEMD runs partially contributed to our ability to cover a wider frequency range and longer chains than previous studies (which did not go over three decades and did not attempt entangled chains).\par
Finally, as shown in \cref{tab:timescales}, each EMD run of the $N=350$ long-chain case costs $3 \times 10^{8}$ time steps. The total cost of three independent EMD runs at $N=350$, which were used in obtaining its $G(t)$, is comparable to the combined cost of all NEMD runs at different frequencies (one run at each frequency). This arrangement allows us to directly compare these two methods at the same computational cost for this particular chain length.

\subsection{Corrected Rouse Mode Analysis (cRMA)}\label{sec:cRMA}
The Rouse model describes the dynamics of an unentangled polymer melt without the topological constraints imposed by other surrounding chains. It describes the relaxation of the polymer melt with a mean-field approach in which effects of surrounding chains on the dynamics of the probe chain are coarse grained as a continuous viscous medium. The equations of motions for the chain beads can be simplified by projecting the original bead coordinates to a set of mutually orthogonal coordinates known as the Rouse modes\cite{rouse1953theory,Kopf1997}
\begin{gather}\label{eq:rouse_mode}
	\vec{X}_p \equiv
	\begin{dcases}
		\sqrt{\frac{1}{N}}\sum_{n=1}^N \vec{r}_i(t) & (p = 0)\\		
		\sqrt{\frac{2}{N}}\sum_{n=1}^N \vec{r}_i(t)\cos\left(\frac{(i-1/2)p\pi}{N}\right) & (p= 1,2,...)\\	
	\end{dcases}
\end{gather}
where $\vec{r}_i$ denotes the original bead position in Cartesian coordinates and $i$ and $p$ are the indices for the beads and Rouse modes, respectively. The $p=0$ mode is proportional to the center of mass coordinates of the chain. Higher modes, $1 < p \leq N-1$, describe the internal relaxation of sub-chain segments of the size of $N/p$ beads. Orthogonality of Rouse modes means that their relaxation dynamics are independent from one another.
Specifically, the TACF of the $p$-th mode
\begin{equation}
	\left\langle\vec{X}_p(t)\vec{X}_p(0)\right\rangle =
	\left\langle\vec{X}_p^2\right\rangle
	\exp\left(-\frac{t}{\tau_p}\right)
	\label{eq:simple_exponential}
\end{equation}
does not depend on any other mode. Its relaxation time scale $\tau_p$ is related to the relaxation time of the first mode $\tau_1$ (same as the Rouse time $\tau_R$) through  $\tau_p = \tau_1 /p^2$. In practice, $\tau_p$ can be obtained by fitting the TACF of the corresponding Rouse mode from EMD to \cref{eq:simple_exponential}. Once $\tau_p$ is known, the $G(t)$ can be calculated by 
\begin{equation}
	G^\text{Rouse}(t) = \frac{\nu k_B T}{N} \sum_{p=1}^{N}\exp \left(-\frac{2t}{\tau_p} \right)
	\label{eq:G_Rouse}
\end{equation}
where $\nu$ is the number density of the beads.\par
Computation of Rouse modes from \cref{eq:rouse_mode} only requires bead positions $\vec{r_i}$ whose fluctuations during an EMD simulation are negligibly small when compared with stress fluctuations. As such, obtaining $G(t)$ from the Rouse modes using \cref{eq:G_Rouse} is expected to produce much lower statistical uncertainty, implying that accurate results can be obtained with shorter runs. In this study, the same EMD data set from Sec. \ref{sec:EMD} was used for computing Rouse modes.\par 
\citet{vladkov2006linear} tested this idea and noted that, for their very short ($N=10$ and $20$) chains, $G'$ from RMA is very close to the NEMD results, while $G''$ from RMA is substantially lower than NEMD. This deficit was attributed to the non-bonded interactions between beads which are mostly excluded in the mean-field approximation of the surrounding chains. Effects of those interactions are felt at time scales shorter than the internal relaxation times of the polymer conformation $\tau_p$. It is thus possible to extract their contributions directly from the short-time limit of the stress TACF, where statistical accuracy is the highest. \citet{vladkov2006linear} proposed to fit the short time part of the $G(t)$ profile from the GK relation \cref{eq:GK} using
\begin{equation}
	G^\text{early}(t) =A\exp\left(-\frac{t}{\tau_A}\right)
		\cos\left(\Omega t\right)
		+ B \exp\left(-\frac{t}{\tau_B}\right)
	\label{eq:stress_fit}
\end{equation}
where $A$, $B$, $\tau_A$, $\tau_B$, and $\Omega$ are fitting parameters. \par
The full $G(t)$ expression for the cRMA approach is then 
\begin{gather}
	G(t) =
	\begin{dcases}
		G^\text{early}(t) & t\leq \tau^*\\
		G^\text{Rouse}(t) & t> \tau^*
	\end{dcases}
\end{gather}
with $G^\text{early}(t)$ and $G^\text{Rouse}(t)$ given by \cref{eq:stress_fit} and \cref{eq:G_Rouse}, respectively. The cut-off time $\tau^*=0.44$ was empirically chosen in this study so that $G^\text{early}(t)$ and $G^\text{Rouse}(t)$ connect continuously. It also sets the upper bound of the GK $G(t)$ data used for parameterizing \cref{eq:stress_fit}. To obtain $G'$ and $G''$, the integrals of \cref{eq:Gprime} and \cref{eq:Gdprime} were correspondingly evaluated as summations of two segments. The first segment integrates from $t=0$ to $\tau^*$ using $G^\text{early}(t)$. For both unentangled ($N=25$ and $50$) cases, the obtained $\tau_A\approx\tau_B\approx 0.1$ are much smaller than $\tau^*$ -- $G^\text{early}(t)$ is vanishingly small at $t>\tau^*$. We thus approximately used integration from $0$ to $\infty$ instead which can be evaluated analytically to give \cite{vladkov2006linear}
\begin{align}
	G^{\prime, \text{early}} &{}= \frac{A}{2} \left(\frac{\omega(\omega + \Omega){\tau_A}^2}{1+{(\omega+\Omega)}^2{\tau_A}^2} + \frac{\omega(\omega -\Omega){\tau_A}^2}{1+{(\omega-\Omega)}^2{\tau_A}^2} \right) \notag \\ 
	&{}+ B\frac{\omega ^2 {\tau_B}^2}{1+\omega ^2 {\tau_B}^2}
\end{align}
\begin{align}
	G^{\prime\prime, \text{early}} &{}=\frac{A \omega \tau_A}{2} \left(\frac{1}{1+{(\omega + \Omega)}^2 {\tau_A}^2}+\frac{1}{1+{(\omega -\Omega)}^2 {\tau_A}^2} \right) \notag \\
	&{} + B \frac{\omega \tau_B}{1+\omega ^2 {\tau_B}^2}.
\end{align}
The second segment integrates from $\tau^*$ to $\infty$ using $G^\text{Rouse}(t)$, which is evaluated numerically using the procedure of Appendix \ref{appendix_A}.
\section{Results and Discussion}\label{label:Results}
In this section, we first present the simulation outputs and data processing for each of the EMD (GK), NEMD, and cRMA approaches (Sec. \ref{sec:results_emd} to \ref{sec:results_cRMA}). $G'$ and $G''$ from these approaches are then compared in Sec. \ref{sec:methods_compared}. Uncertainty and computational cost considerations are discussed in Sec. \ref{sec:cost_disc}
\subsection{EMD Results}\label{sec:results_emd}
\begin{figure}
	\centering
	\includegraphics[width=3.5in, trim=0 0 0 0, clip]{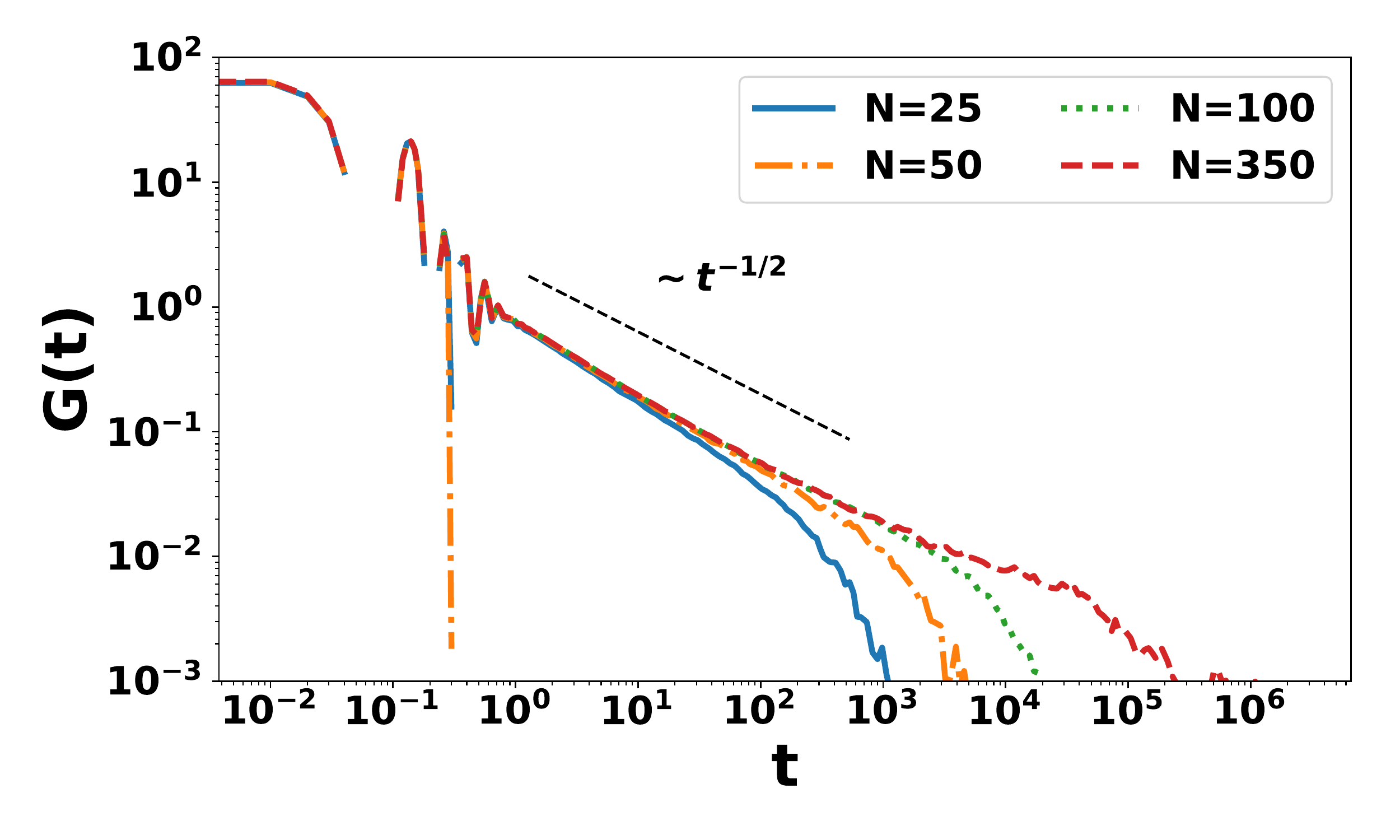}
	\caption{Shear stress relaxation modulus $G(t)$ of varying chain length using EMD results and the GK relation.}
	\label{fig:Gt_plot}
\end{figure}

The $G(t)$ profiles from EMD using the GK relation are shown in \cref{fig:Gt_plot}. It can be seen that the hierarchical averaging in the multi-tau correlator method has effectively erased noise in the $G(t)$ for nearly the whole time range of interest. At early times, the curves all collapse on one another. The wild oscillations at early times come from bond fluctuations and the curves appear broken because negative values are not shown in the logarithmic scale. This is followed in all cases by a $t^{-1/2}$ scaling regime.
\RevisedText{The $t^{-1/2}$ scaling is predicted from the Rouse model for the stress relaxation of unentangled polymer chains.
For entangled chains, the scaling is expected in sub-entanglement scales.
Interestingly, the same $t^{-1/2}$ scaling was also recently reported for simple random bead-spring networks by \citet{Milkus2017}, which may suggest its more general origin in disordered materials.}

The curves separate at later times. The shorter chains ($N= 25$ and $50$) decay exponentially after their respective Rouse times.
For longer chains, however, the relaxation is prolonged as a result of entanglement. Departure from the Rouse relaxation is most visible for the longest $N=350$ case. (Departure from the $t^{-1/2}$ Rouse scaling for the $N=350$ case was confirmed in our earlier study\cite{adeyemi2021dynamics}). At $N=350$, the chains are not yet deeply entangled and thus $G(t)$ does not develop a full-fledged stress plateau which is not expected until $N\gg N_\text{e}$\cite{likhtman2010comment,hsu2016static}.
\subsection{NEMD Results}\label{sec:results_nemd}
\begin{figure*}
	\centering
	\begin{subfigure}[b]{0.49\textwidth}
		\centering
		\includegraphics[width=\textwidth]{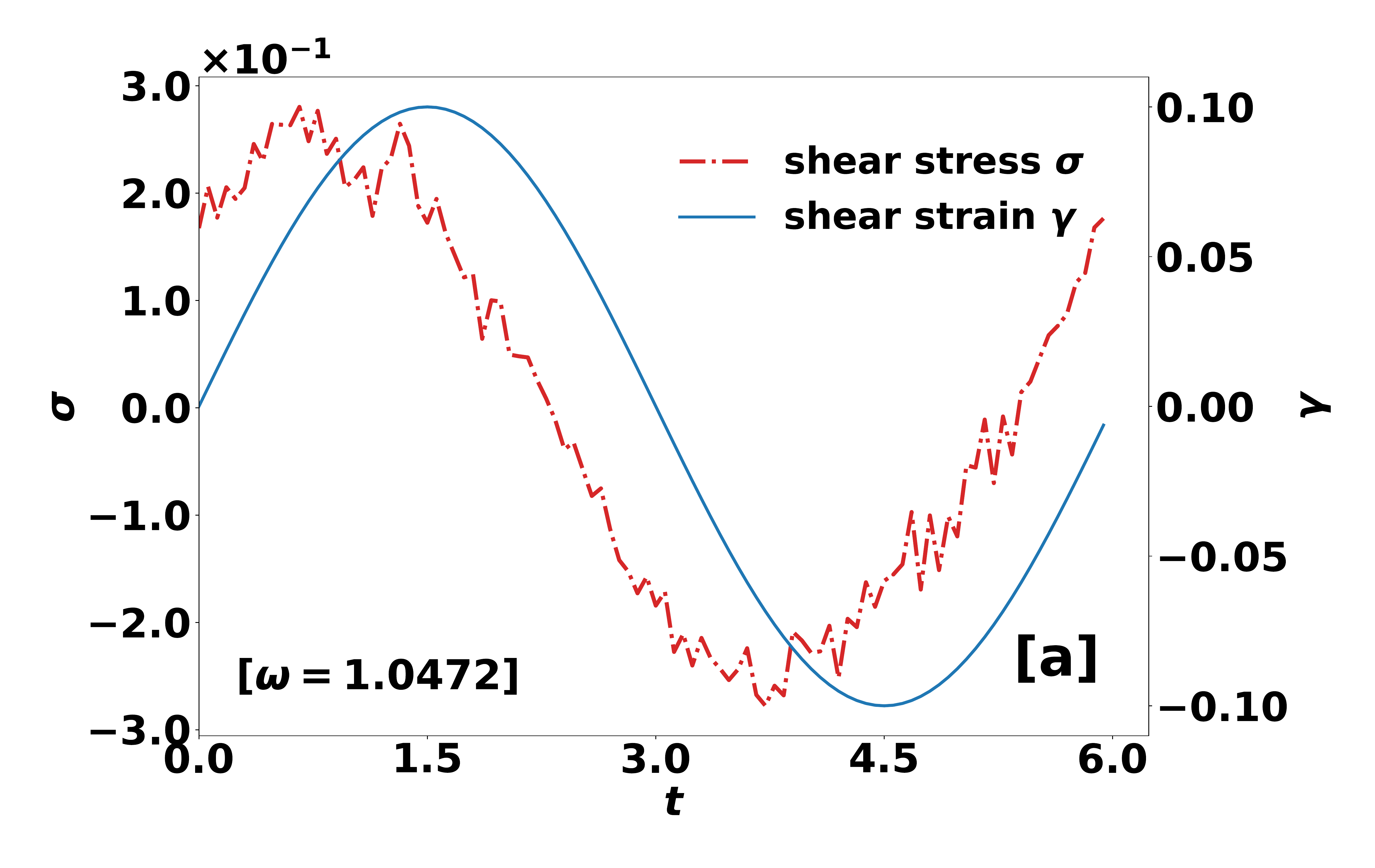}
		\phantomsubcaption
		\label{fig:6.0_plot}
	\end{subfigure}
	\begin{subfigure}[b]{0.49\textwidth}
		\centering
		\includegraphics[width=\textwidth]{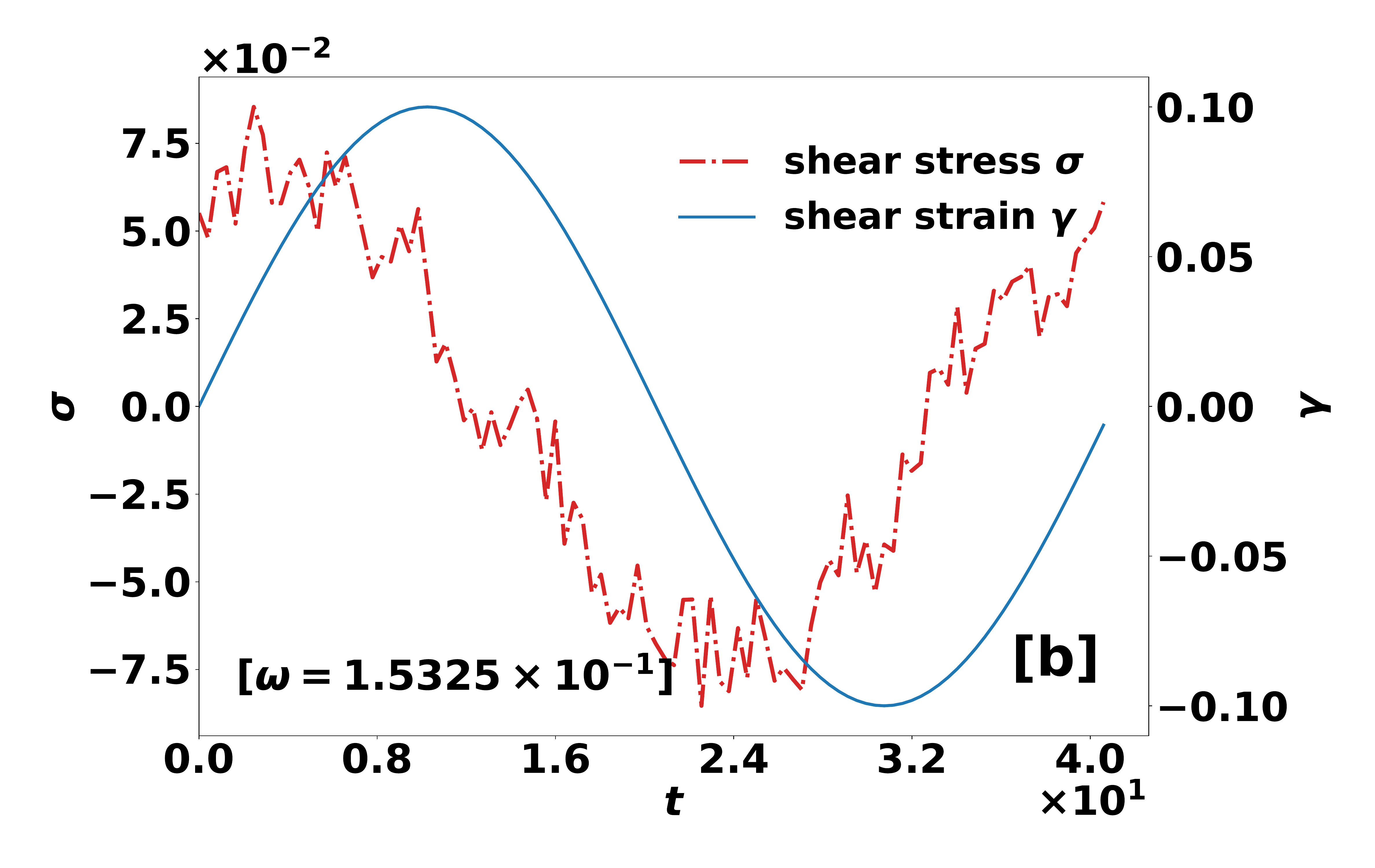}
		\phantomsubcaption
		\label{fig:41.0_plot}
	\end{subfigure}
	\begin{subfigure}[b]{0.49\textwidth}
		\centering
		\includegraphics[width=\textwidth]{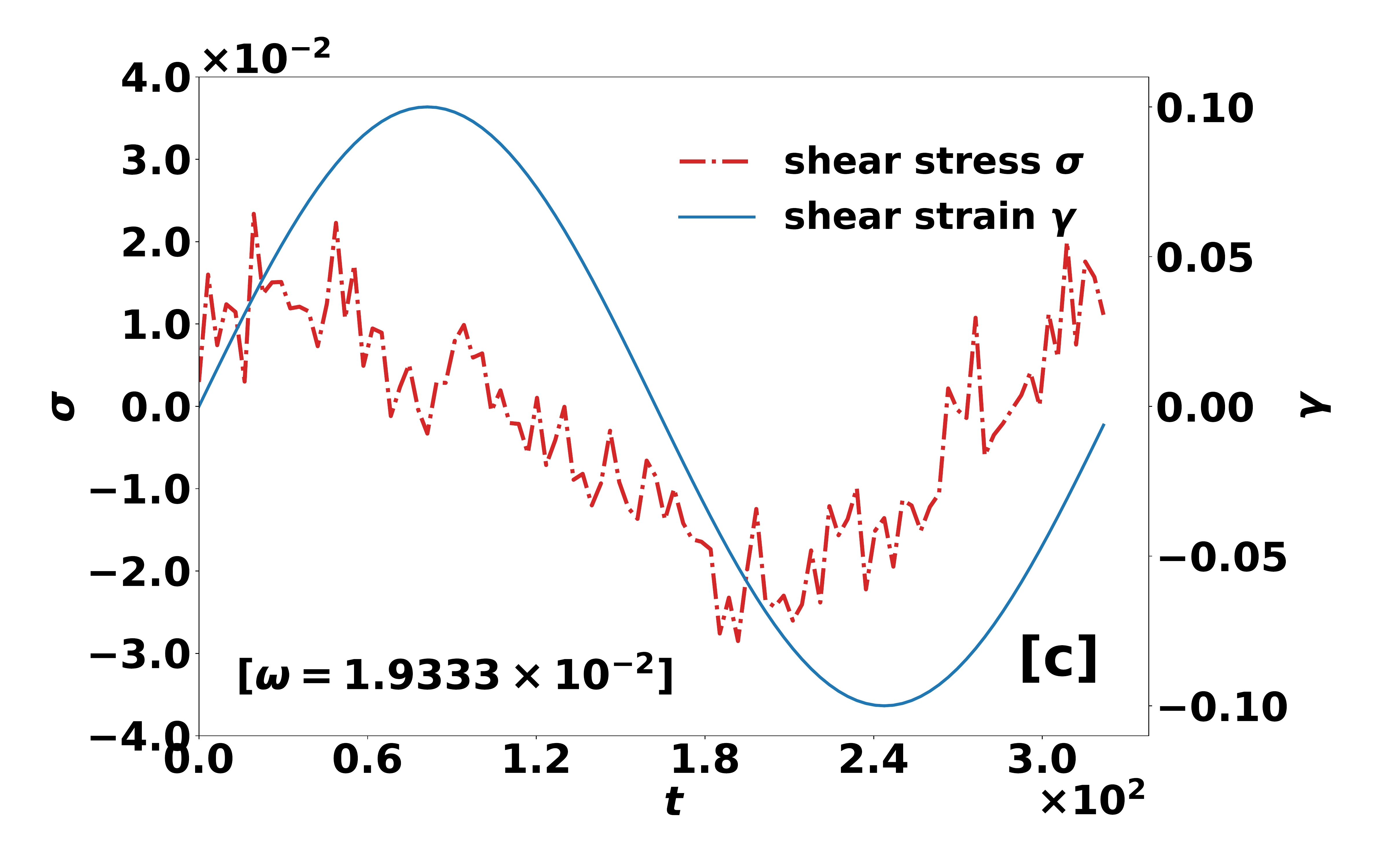}
		\phantomsubcaption
		\label{fig:325.0_plot}
	\end{subfigure}
	\begin{subfigure}[b]{0.49\textwidth}
		\centering
		\includegraphics[width=\textwidth]{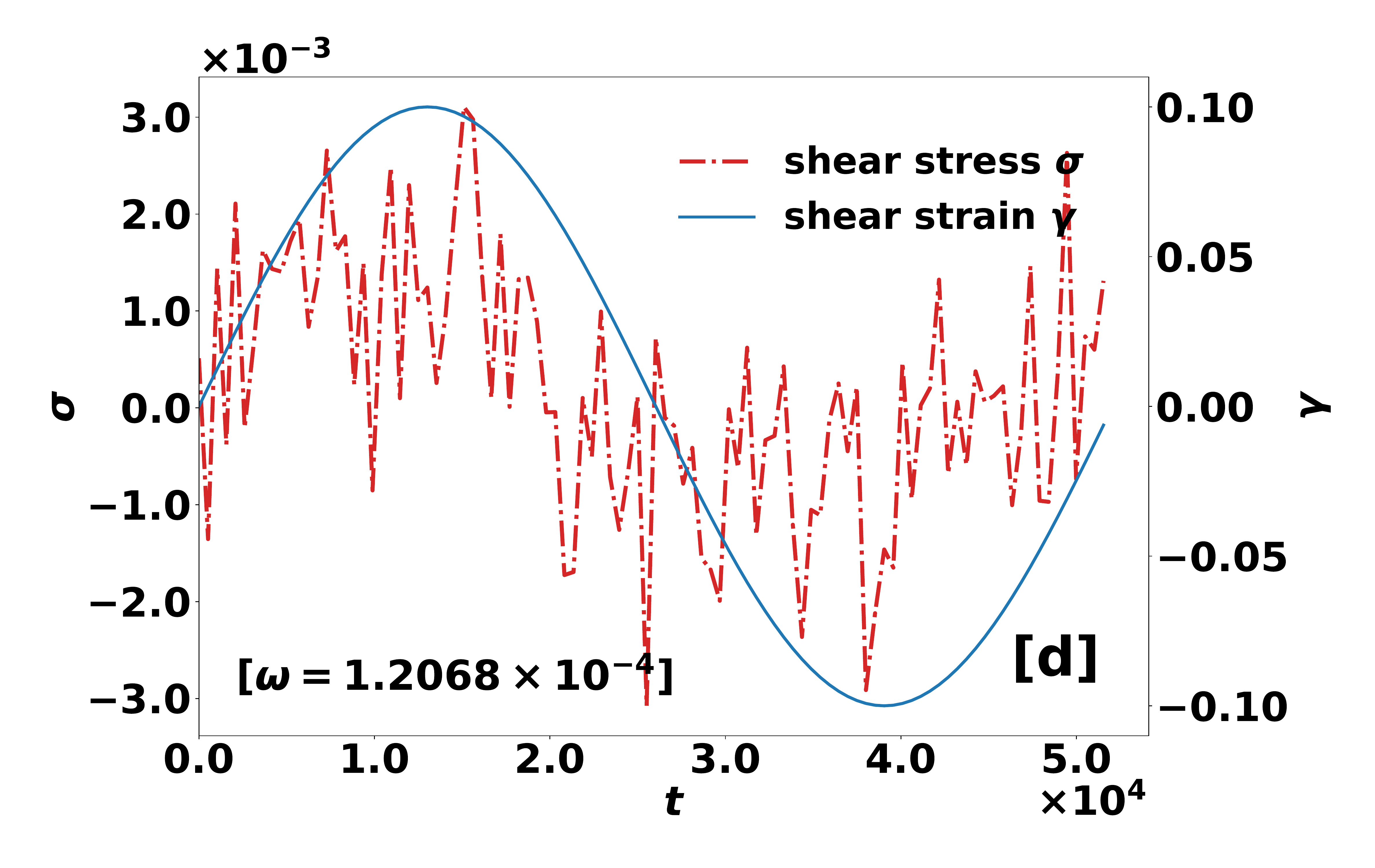}
		\phantomsubcaption
		\label{fig:52065.0_plot}
	\end{subfigure}
	\label{fig:nemd_ind}	
	\caption{Stress-strain time series of a typical converged cycle at each frequency for $N=350$ (a) $\omega = 1.0472$ (b) $\omega = 1.5325 \times 10^{-1}$ (c) $\omega = 1.9333 \times 10^{-2}$ (d) $\omega = 1.2068 \times 10^{-4}$ (all using $\gamma_0=0.1$). The stress signal at each frequency has been pre-averaged over a window size of $1/100$ of the cycle.}
\end{figure*}
For a sinusoidal strain deformation that is small enough to still be in the linear regime, it is expected that the resulting stress is equally sinusoidal and oscillates with the same frequency as the strain but with a phase shift reflecting the viscoelasticity of the material.
The first thing to check is thus whether the resulting stress is indeed oscillating with the same frequency. \Crefrange{fig:6.0_plot}{fig:52065.0_plot} show the stress and strain time series for different frequencies for the longest chain $N = 350$. It can be seen that the stress indeed oscillates at the same frequency as the strain with a notable phase lead. Despite the pre-averaging treatment mentioned in Sec. \ref{sec:NEMD}, the resulting stress signal still contains substantial noise. As frequency decreases, the stress magnitude is lower and the noise-to-stress ratio is higher.\par
\begin{figure}
	\centering
	\includegraphics[width=3.5in, trim=0 0 0 0, clip]{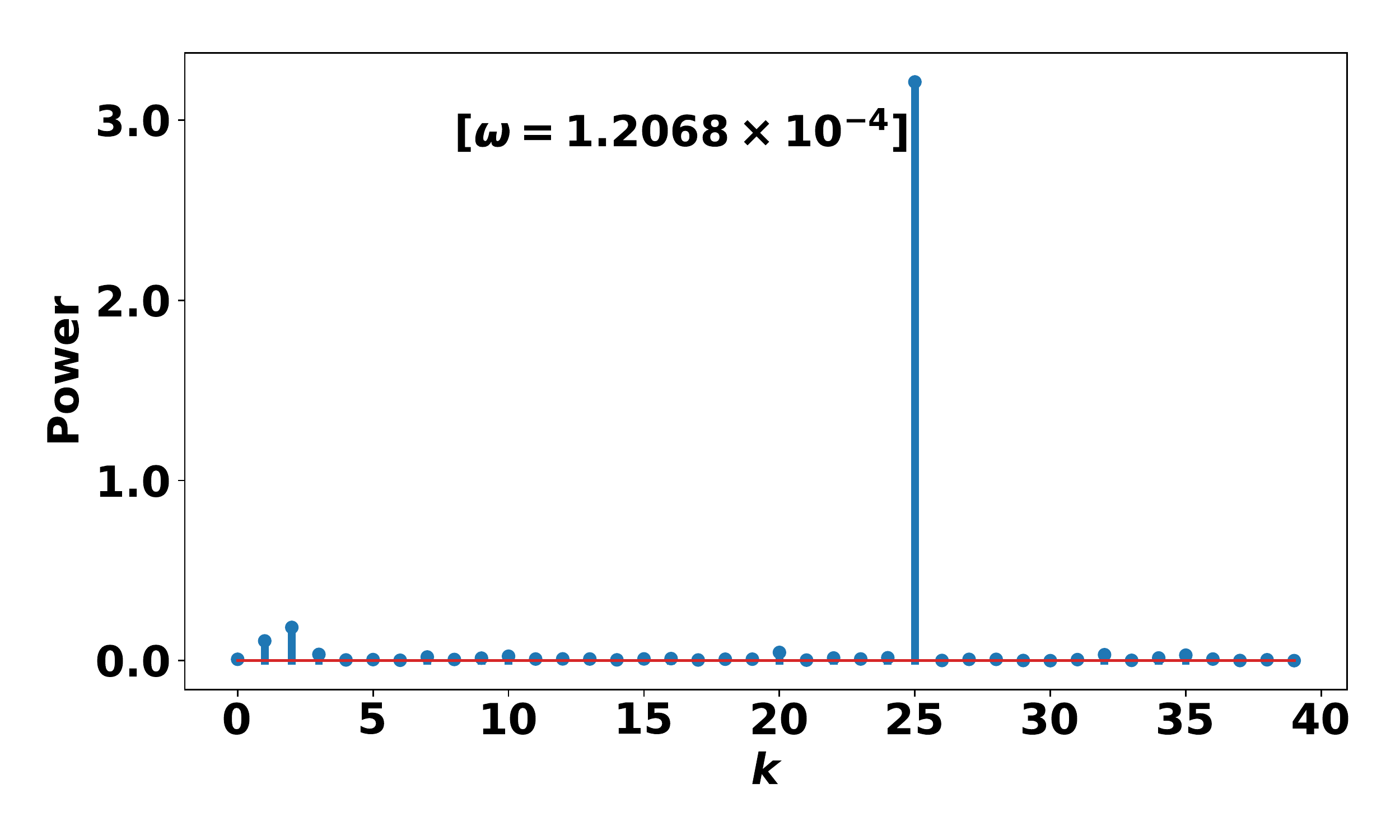}
	\caption{Power spectrum of the stress signal for input frequency $\omega = 1.2068 \times 10^{-4}$. The first 40 modes are shown.}
	\label{fig:52065.0_stem_plot}
\end{figure}

To demonstrate the effectiveness of DFT in extracting the dominant mode for $G'$ and $G''$, we take the lowest frequency ($\omega = 1.2068 \times 10^{-4}$) case in \cref{fig:52065.0_plot} as an example, where the noise level appears comparable to the amplitude of the primary oscillation. \Cref{fig:52065.0_stem_plot} shows the power spectrum of its stress time series (all 25 cycles included in the statistics), as defined by
\begin{equation}
	P_k = |\tilde{c}_k|^2
\end{equation}
for the leading wavenumbers. Here, $P_k$ is the power associated with the $k$-th mode and $\tilde{c}_k$ is its complex Fourier coefficient. Since the whole time series contains 25 cycles, the primary mode is expected at $k=25$. The power magnitude at $k=25$ is indeed distinctly higher than the rest of the spectrum (despite the large noise seen in \cref{fig:52065.0_plot}). Its real and imaginary parts are used to calculate $G''$ and $G'$ respectively, according to \cref{eq:gg_fourier}.
An equally high peak is expected at the $(N_t-25)$-th mode. ($N_t$ is the total number of points in the time series). Its Fourier coefficient is simply the complex conjugate of the $k=25$ mode.
\par

\begin{figure}
	\centering
	\begin{subfigure}[b]{\linewidth}
		\centering
		\includegraphics[width=\linewidth]{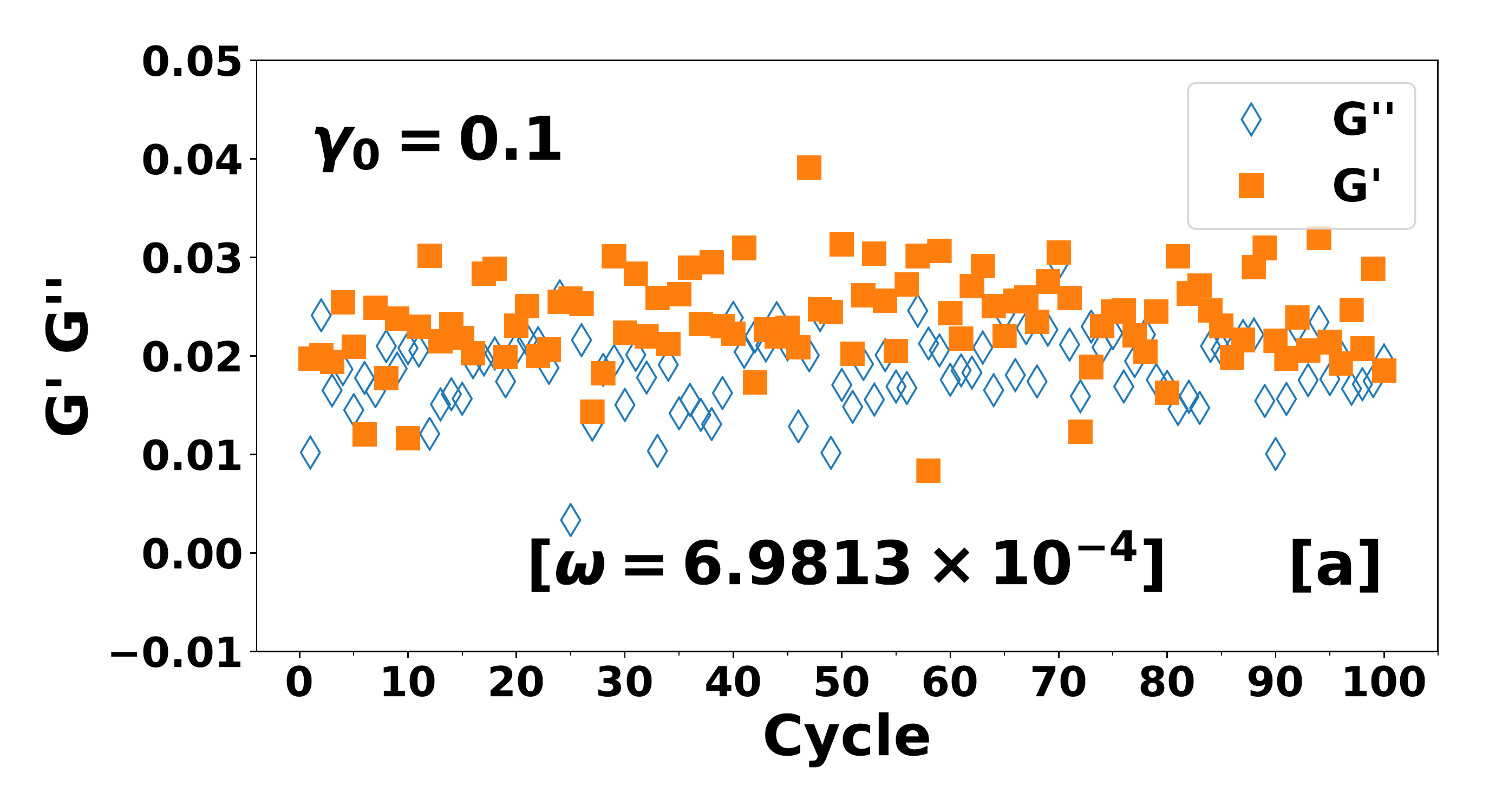}
		\phantomsubcaption
		\label{fig:9000_350_0.1_Gdpr_blocks}
	\end{subfigure}
	\begin{subfigure}[b]{\linewidth}
		\centering
		\includegraphics[width=\linewidth]{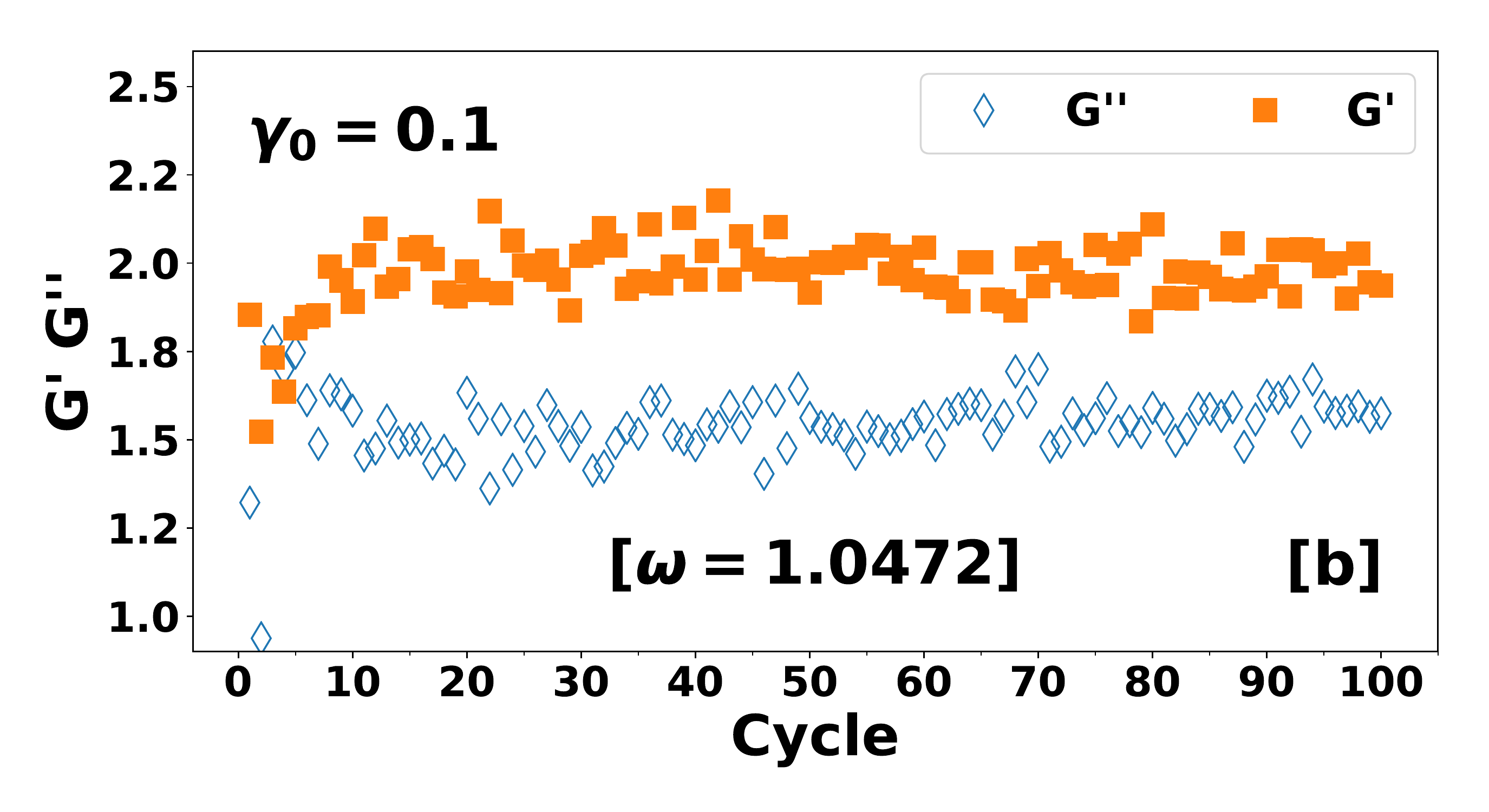}
		\phantomsubcaption
		\label{fig:6_350_0.10_Gdpr_blocks}
	\end{subfigure}
	\begin{subfigure}[b]{\linewidth}
		\centering
		\includegraphics[width=\linewidth]{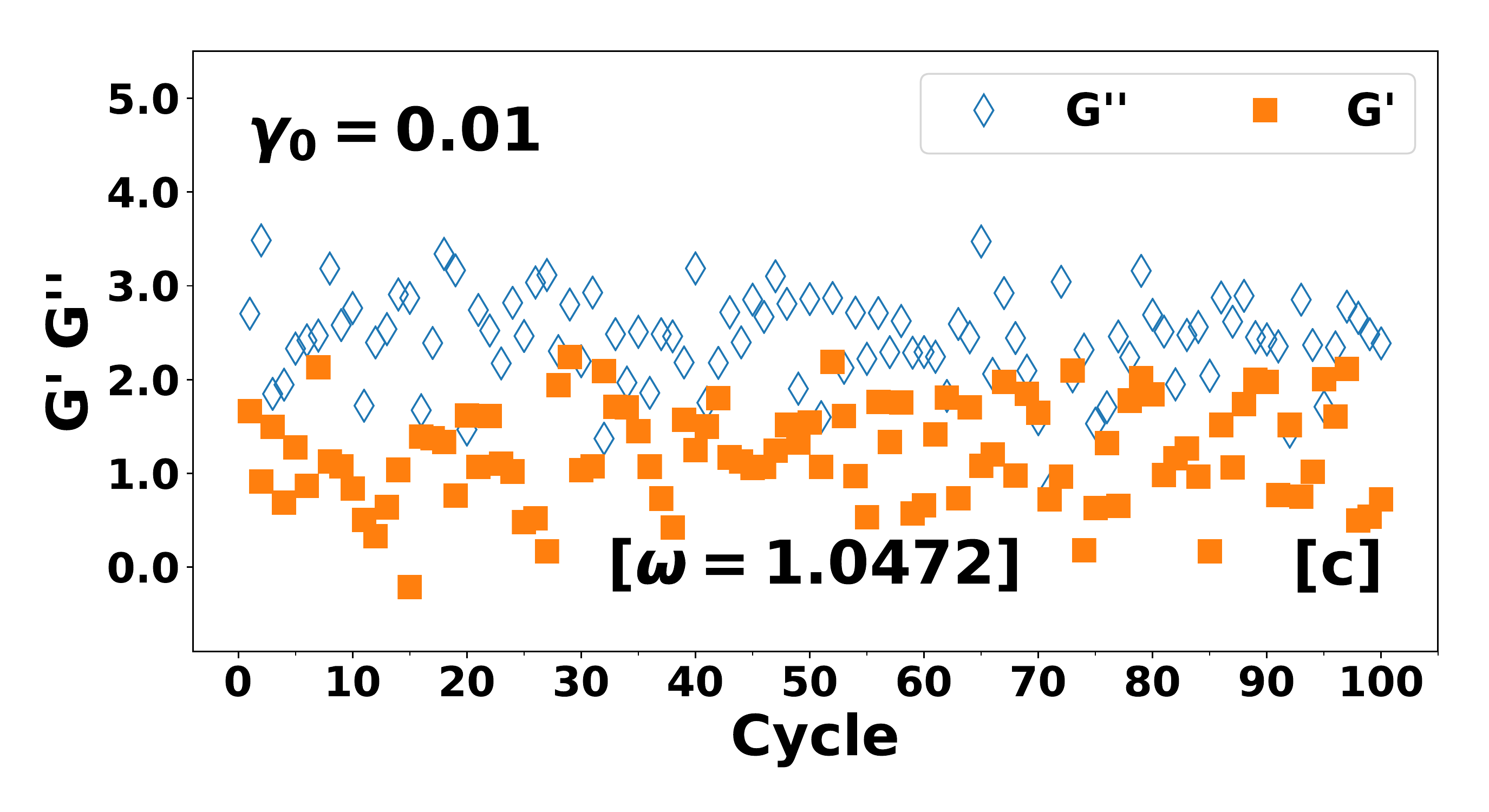}
		\phantomsubcaption
		\label{fig:6_350_0.01_Gdpr_blocks}
	\end{subfigure}	
	\caption{G' and G'' calculated from individual cycles for $N=350$: (a) $\gamma_0=0.1$ and $\omega = 6.9813 \times 10^{-4}$; (b) $\gamma_0=0.1$ and $\omega = 1.0472$; (c) $\gamma_0=0.01$ and $\omega = 1.0472$.}
	\label{fig:GG_cycles}
\end{figure}

Data in \cref{fig:52065.0_stem_plot} come from 25 cycles with a maximum strain amplitude $\gamma_0=0.1$. The $\gamma_0$ value was chosen based on previous studies which reported that $0.1$ falls well within the linear regime where the complex moduli do not depend on the strain magnitude \cite{cifre2004linear,vladkov2006linear}. Our chosen $N_\text{cycle}=25$ is, however, substantially lower than those same previous studies (which used $100$ to $200$ cycles). To justify this choice, we divide the whole time series into individual cycles. Applying DFT to each cycle renders its own $G'$ and $G''$ values. \Cref{fig:GG_cycles} shows these single-cycle $G'$ and $G''$ values for extended $100$-cycle simulation runs.
For $\gamma_0=0.1$, at low frequency (\cref{fig:9000_350_0.1_Gdpr_blocks}), results from all cycles fluctuate around common mean values, but at high frequency (\cref{fig:6_350_0.10_Gdpr_blocks}), the results do not converge statistically until a transient period is passed. The transient period seems to depend on both frequency and chain length. The particular case in \cref{fig:6_350_0.10_Gdpr_blocks} shows a transient period lasting for $\sim 20$ cycles but transient periods as long as $\sim 40$ cycles were observed in other cases. As such, when reporting data from these high-frequency cases, the transient period must be discarded and the following 25 cycles in the converged regime should be used.\par
The computational overhead introduced by those extra transient cycles is small since they only affect the least expensive, high-frequency regime. However, the fact that, starting from the equilibrium state, it requires a number of cycles for the system to converge to steady oscillation suggests that perturbation to the equilibrium is substantial --i.e., the oscillatory shear may no longer belong to the linear regime. Since the linear and non-linear regimes are separated based on the Weissenberg number $\mathrm{Wi}\equiv\tau_\text{relax}\gamma_0\omega$ ($\tau_\text{relax}$ is the polymer relaxation time), transition to the non-linear regime occurs at lower $\gamma_0$ for higher $\omega$. Indeed, for the same frequency and chain length, if we reduce $\gamma_0$ to $0.01$ (\cref{fig:6_350_0.01_Gdpr_blocks}), the transient period is no longer observed. This effect of strain magnitude will be further discussed below when we compare $G'$ and $G''$ results. \par

\begin{figure}
	\centering
	\begin{subfigure}[b]{0.48\textwidth}
		\centering
		\includegraphics[width=\textwidth]{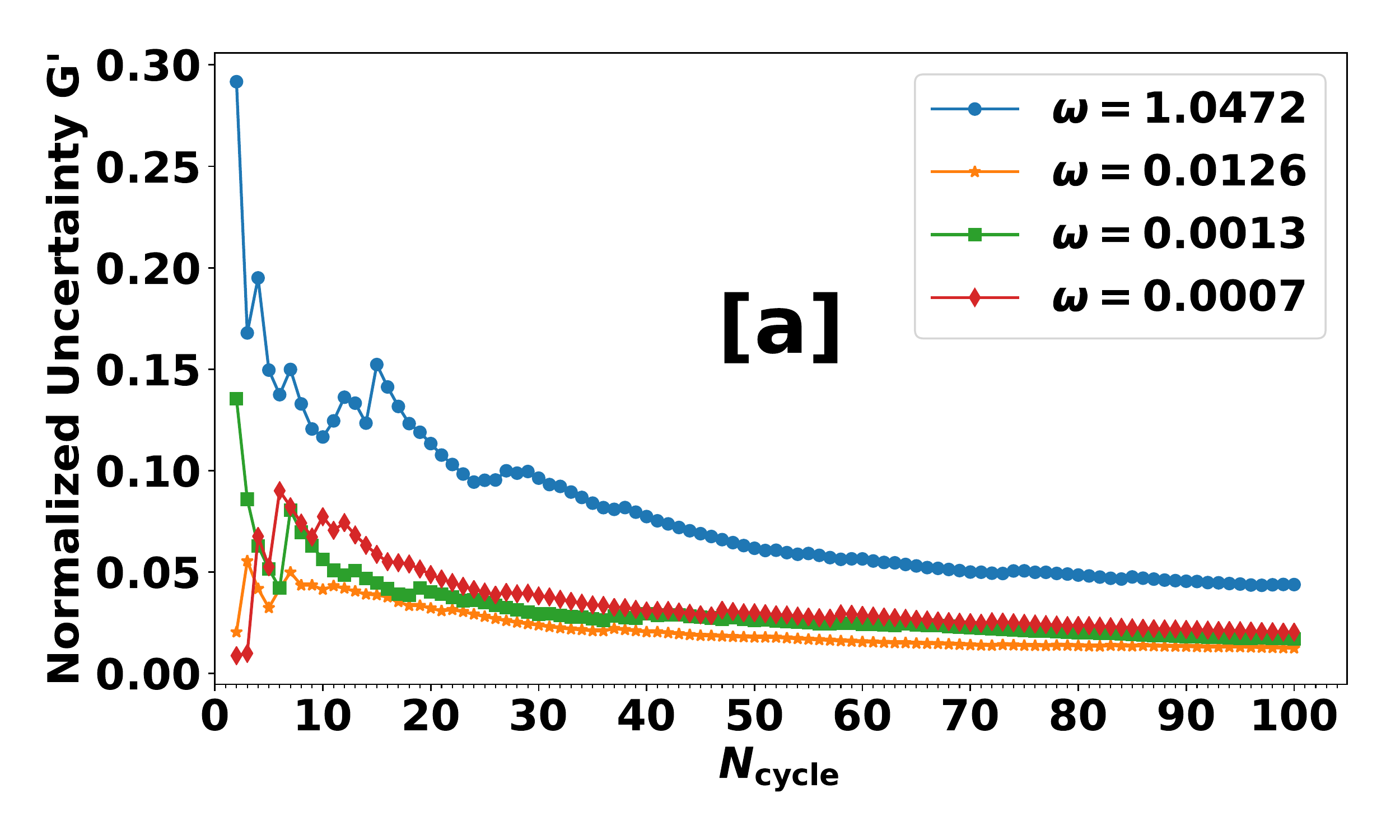}
		\phantomsubcaption{}
		\label{fig:1_gpr_unc}
	\end{subfigure}
	\begin{subfigure}[b]{0.48\textwidth}
		\centering
		\includegraphics[width=\textwidth]{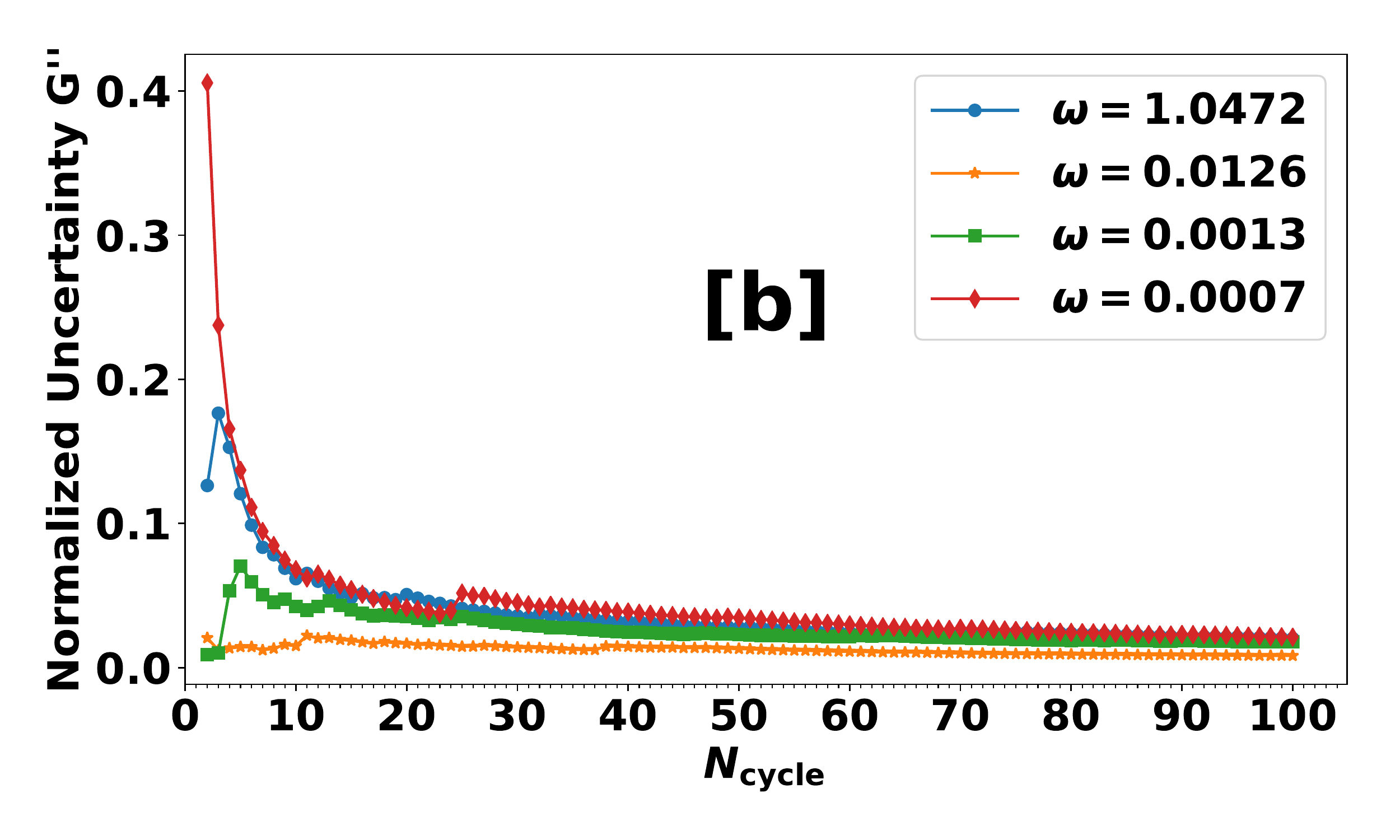}
		\phantomsubcaption{}
		\label{fig:1_gdpr_unc}
	\end{subfigure}
	\caption{Uncertainty of (a) $G'$ and (b) $G''$ with increasing number of cycles included in the time series, normalized by the estimated $G'$ and $G''$  values ($N=350$).}
	\label{fig:gg_uncertainty}
\end{figure}

\Cref{fig:gg_uncertainty} shows the effects of $N_\text{cycle}$ on the normalized uncertainty in the results. The uncertainty of, e.g., $N_\text{cycle}=10$, was estimated by the standard error of the 10 individual measurements coming from each cycle, which was then normalized by the overall measurement from all 10 cycles combined. There is an initial rapid decrease in uncertainty at the small $N_\text{cycle}$ end but as more cycles are included in the statistics, the marginal gain of increasing the simulation length diminishes. \Cref{fig:gg_uncertainty} only shows the $N=350$ case but the observation is similar for other chain lengths. In all cases, the uncertainty becomes reasonably small for $N_\text{cycle} \geq 25$. We have repeated the analysis with larger block size --i.e. instead of using single cycles, we used every two or every five cycles as an individual measurement and still arrived at the same conclusion.

\begin{figure}
	\centering
	\includegraphics[width=3.5in, trim=0 0 0 0, clip]{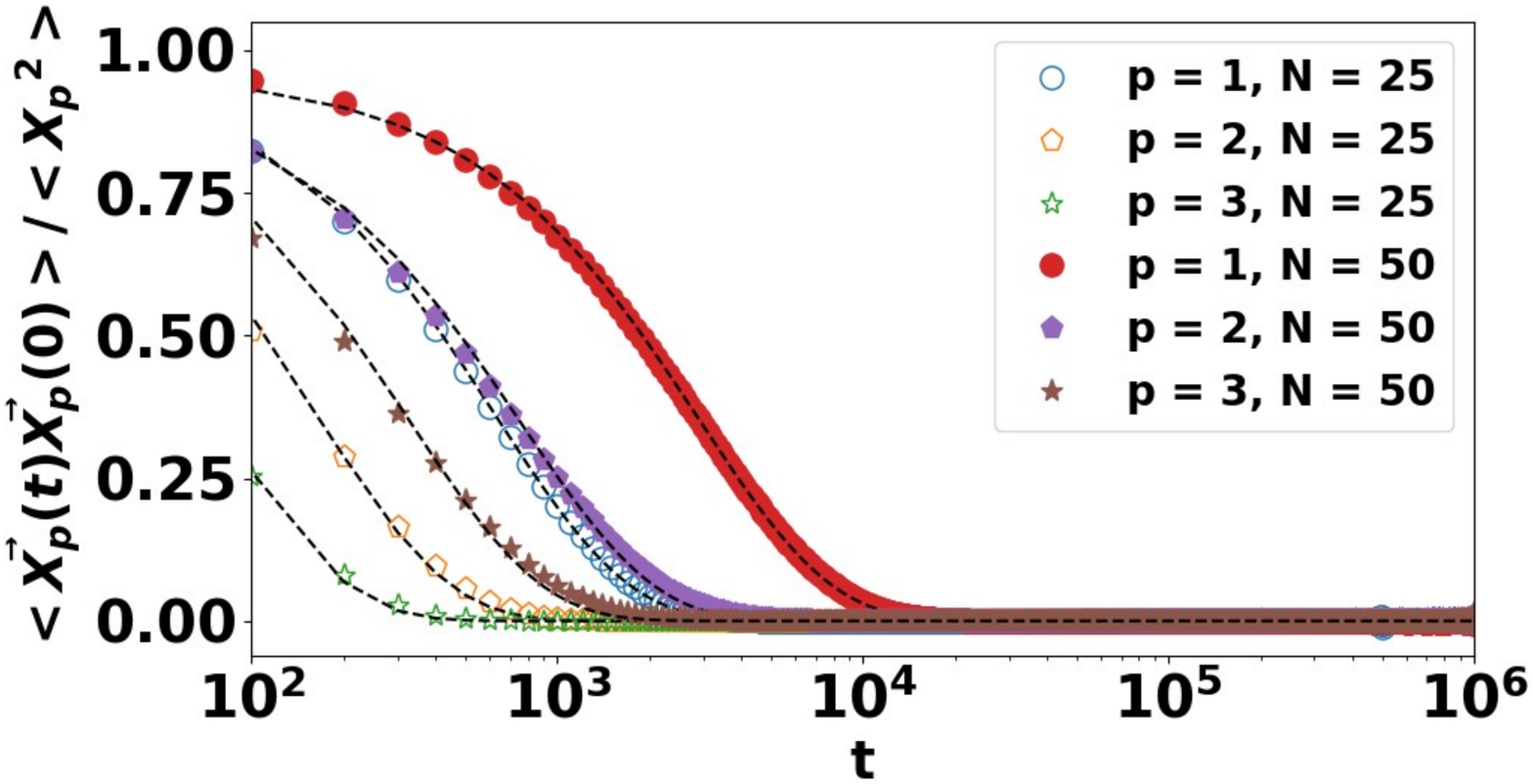}
	\caption{Relaxation of $p = 1,2,$ and $3$ Rouse modes for $N = 25$ (filled markers) and $N = 50$ (empty markers). Lines represent fitted regression lines using the simple exponential relaxation function of \cref{eq:simple_exponential}.}
	\label{fig:Rouse_relaxation}
\end{figure}
\subsection{cRMA Results}\label{sec:results_cRMA}
The cRMA approach is only applicable to shorter unentangled chains. \Cref{fig:Rouse_relaxation} shows the TACFs of the first 3 modes for $N =25$ and $50$ calculated using \cref{eq:simple_exponential} from EMD runs. The profiles are normalized with $\langle \boldsymbol{X}_p^2\rangle$ and thus all start at 1 at the $t=0$ limit, which is not shown in \cref{fig:Rouse_relaxation} due to the logarithmic scale used. Smooth exponential decay can be readily seen in all profiles. One may note that the $p=1$ mode of $N=25$ nearly overlaps with the $p=2$ mode of $N=50$. This is because the $p=2$ mode describes the relaxation of a sub-chain segment with half of the total chain length, which, in the case of $N=50$, happens to be 25 monomers. Fitting the TACF profiles to \cref{eq:simple_exponential} yields the relaxation times for the modes $\tau_p$. For the same chain length, the Rouse model prediction of $\tau_p=\tau_1/p^2$ is approximately held: e.g., for $N=50$, $\tau_1 =2906.25 $, $\tau_2 =761.03$, and $\tau_3 =325.78$. \par
\begin{figure}
	\centering
	\begin{subfigure}[b]{0.49\textwidth}
		\centering
		\includegraphics[width=\textwidth]{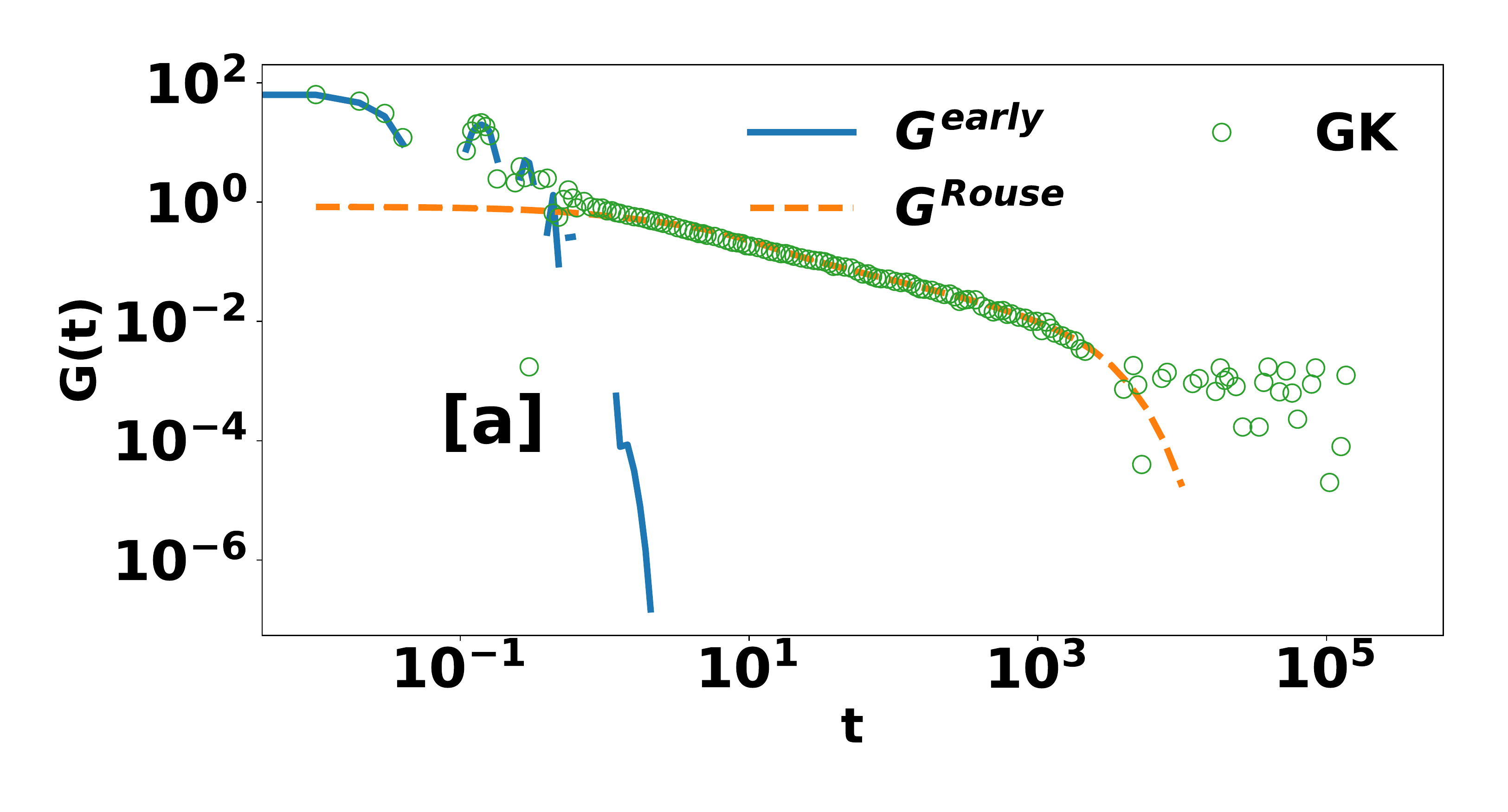}
		\phantomsubcaption
		\label{fig:50_Gplot_Rouse}
	\end{subfigure}
	\begin{subfigure}[b]{0.49\textwidth}
		\centering
		\includegraphics[width=\textwidth]{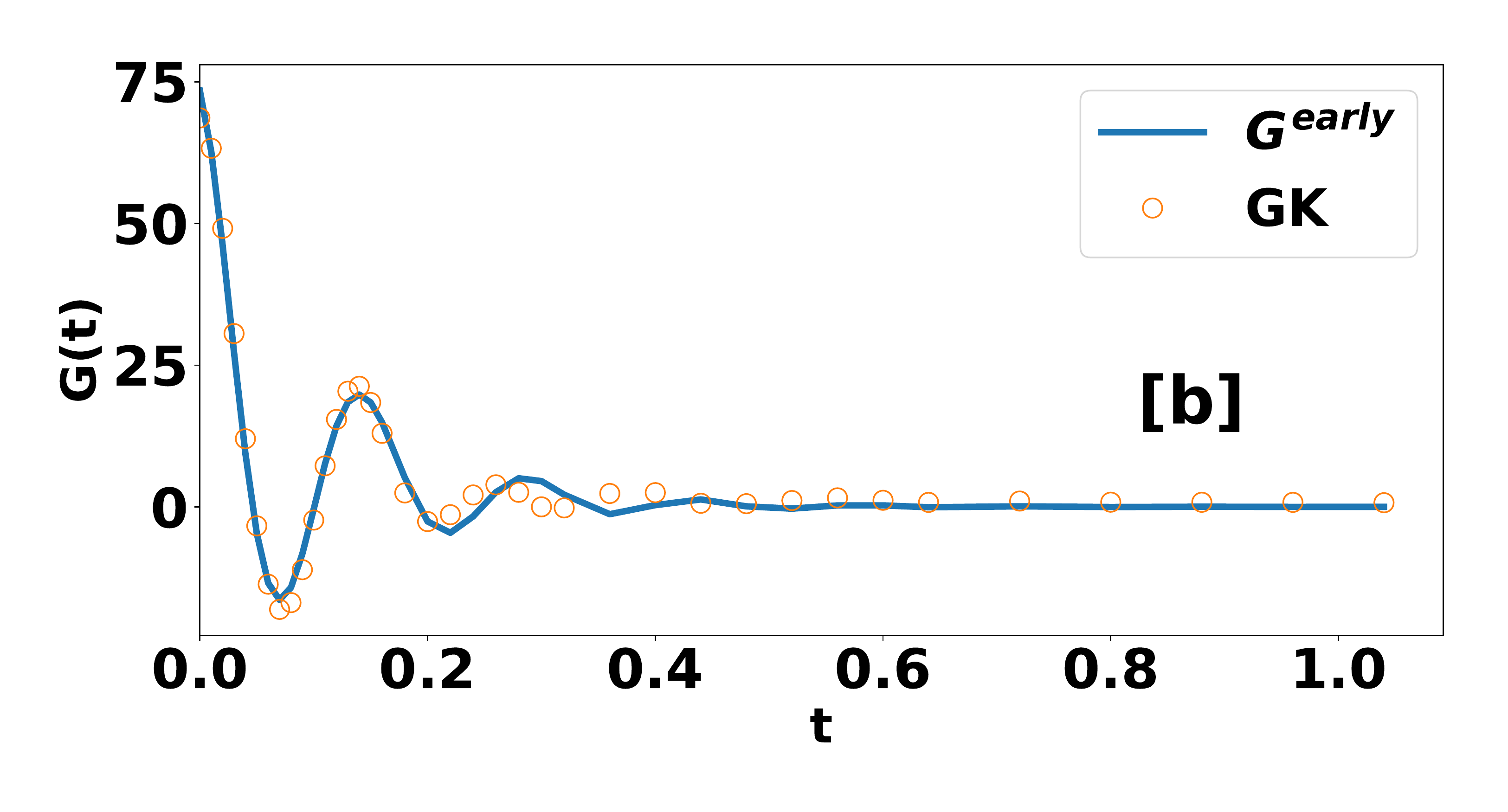}
		\phantomsubcaption
		\label{fig:50_early_times_Rouse}
	\end{subfigure}
	\caption{Relaxation modulus from the Rouse model $G^\text{Rouse}(t)$ compared with that from the Green-Kubo relation $G(t)$ (both using EMD data for $N=50$); $G^\text{early}(t)$ is a fit to the short-time part of $G(t)$: (a) full view; (b) enlarged view of the short-time regime.}
	\label{fig:early_rouse}
\end{figure}

The obtained $\tau_p$ values were used to compute the Rouse-model prediction $G^\text{Rouse}(t)$ per \cref{eq:G_Rouse} which is then plotted in \cref{fig:early_rouse} along with the GK result. It is clear that the Rouse model accurately captures the GK result for over three decades. Discrepancy is noted at $t\gtrsim\mathcal{O}(10^4)$ where the stress has nearly vanished and the GK result is laden with noise. At the short-time end ($t\lesssim\mathcal{O}(1)$), $G^\text{Rouse}(t)$ is significantly lower than the GK $G(t)$ profile. This deficit is attributed to the the bead-bead non-bonded interactions not fully captured by the Rouse model and will result in an underestimate of $G''$ especially at the high frequency regime\cite{vladkov2006linear}. A correction is introduced by fitting the short-time part of $G(t)$ to \cref{eq:stress_fit}. The resulting $G^\text{early}(t)$ captures the short-time $G(t)$ profile well (\cref{fig:50_early_times_Rouse}) but decays quickly after $t\approx 0.4$. Calculation of $G'$ and $G''$ in the cRMA approach combines $G^\text{early}(t)$ and $G^\text{Rouse}(t)$ according to the procedure in Sec. \ref{sec:cRMA}.
\subsection{Comparison of Methods}\label{sec:methods_compared}
\begin{figure}
	\centering
	\begin{subfigure}[b]{0.48\textwidth}
		\centering
		\includegraphics[width=\textwidth]{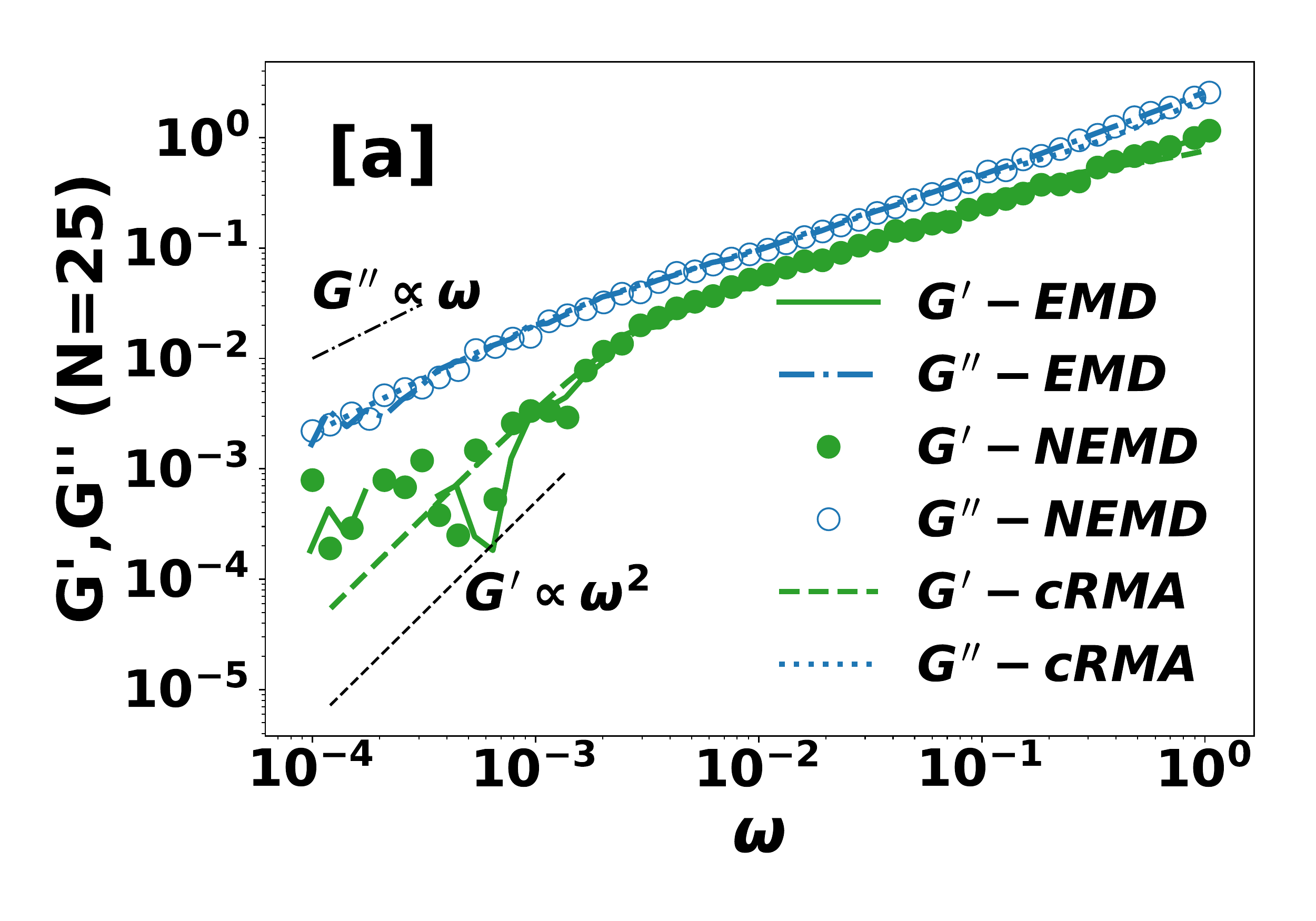}
		\phantomsubcaption
		\label{fig:25_Gpr_emd_nemd_fit}
	\end{subfigure}
	\begin{subfigure}[b]{0.48\textwidth}
		\centering
		\includegraphics[width=\textwidth]{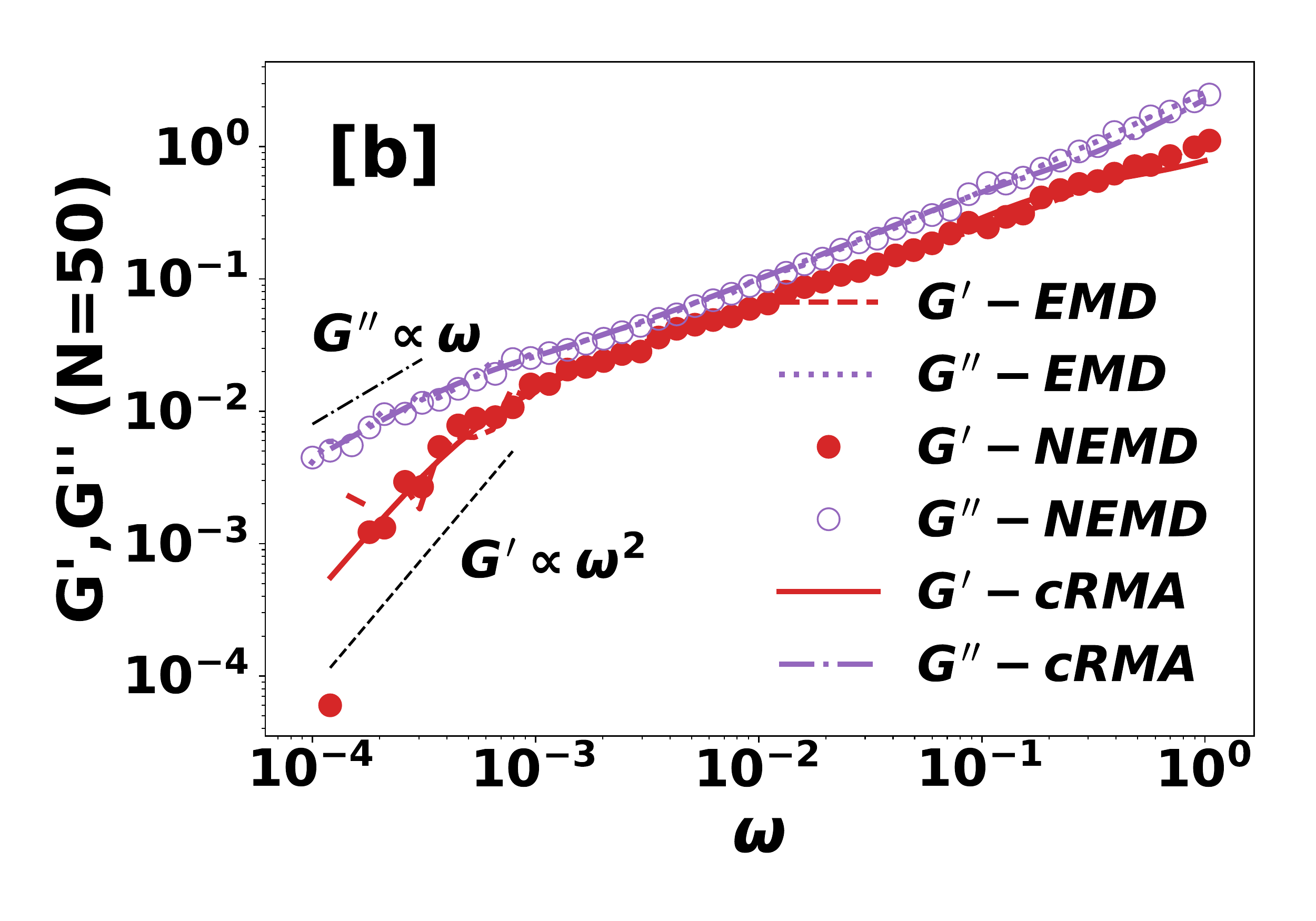}
		\phantomsubcaption
		\label{fig:50_Gpr_emd_nemd_fit}
	\end{subfigure}
	\caption{$G'$ and $G''$ using EMD/GK, NEMD, and cRMA methods for (a) $N=25$ and (b) $N=50$ ($\gamma_0=0.01$ is used for $\omega\geq 7.2222 \times 10^{-2}$ and $\gamma_0=0.1$ used for lower $\omega$).}
	\label{fig:25_50_GG}
\end{figure}

We turn now to the comparison of the computed $G'$ and $G''$ profiles. We first show the results of the shorter chains $N = 25$ and $50$ in \cref{fig:25_Gpr_emd_nemd_fit} and \cref{fig:50_Gpr_emd_nemd_fit}.
Since both types of chains are well within the unentangled regime~\citep{adeyemi2021dynamics}, a rubbery plateau does not exist in the $G'$ profile. We further observe the Rouse scaling -- $G' \propto \omega^2$ and $G'' \propto \omega$ at the terminal (low $\omega$) frequencies for both chains. The $G''$ values are greater than the $G'$ values at all frequencies. For $N=25$ in \cref{fig:25_Gpr_emd_nemd_fit}, all three methods (EMD/GK, NEMD, cRMA) give nearly equivalent results for intermediate and high frequencies ($10^{-3}$ and above) for both $G'$ and $G''$. The agreement is equally good at the low frequency end for $G''$, but for $G'$, strong fluctuations are found in both the EMD/GK and NEMD results. The high noise-to-signal ratio is most likely due to the low magnitude of $G'$ in that regime, which reflects the quick relaxation of the $N=25$ chains. For $N=50$ in \cref{fig:50_Gpr_emd_nemd_fit}, the results are very similar to $N=25$ except that fluctuations in $G'$ at the low-$\omega$ end appear smaller especially in the NEMD case.\par
Of the three methods, cRMA is least affected by simulation noise and uncertainty. This does not come as much of a surprise because the cRMA method is based on particle coordinates from the EMD simulation and avoids the intrinsically noisy stress calculation, with the only exception of the short-time stress correlation used in the correction term. Comparison with the EMD/GK and NEMD results shows that cRMA also produces reliable results for linear viscoelastic properties. However, its usage is limited to strictly unentangled polymers.\par
\begin{figure}
	\centering
	\includegraphics[width=3.5in, trim=0 0 0 0, clip]{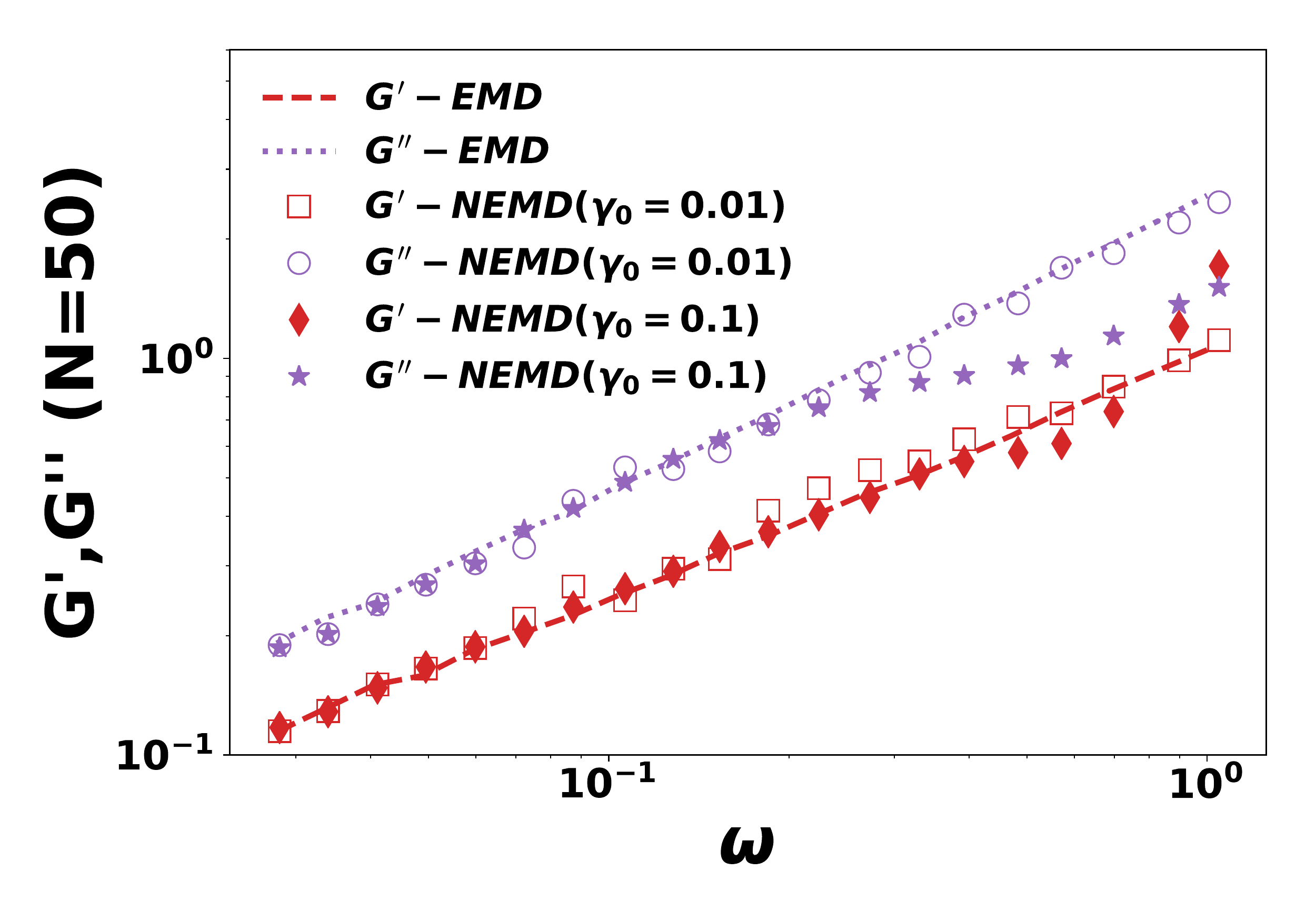}
	\caption{Effects of strain amplitudes on the NEMD results in the high frequency regime ($N=50$).}
	\label{fig:50_Gpr_emd_nemd_fit_kink}
\end{figure}

 
Unless otherwise noted, NEMD results here and below used a standard strain amplitude of $\gamma_0=0.1$, except in the high frequency regime ($\omega\geq 7.2222 \times 10^{-2}$) where $\gamma_0=0.01$ was used. This is because the standard $\gamma_0=0.1$ would yield unreliable results at higher frequencies. \Cref{fig:50_Gpr_emd_nemd_fit_kink} shows the comparison between these two strain amplitudes in the $N=50$ case as an example. The standard $\gamma_0=0.1$ is accurate for frequency up to $\omega = 0.2244$, after which unnatural kinks are found in both profiles, with $G'$ and $G''$ being respectively over- and under-estimated compared with EMD results. Similar behaviors are found in all other chain lengths studied. Reducing $\gamma_0$ to $0.01$ produces results that not only extend smoothly from the $\gamma_0=0.1$ results of lower frequency, but also agree well with EMD results. This corroborates our earlier discussion that for $\gamma_0=0.1$, the flow is no longer in the linear regime at high frequency.\par
\begin{figure}
	\centering
	\includegraphics[width=3.5in, trim=0 0 0 0, clip]{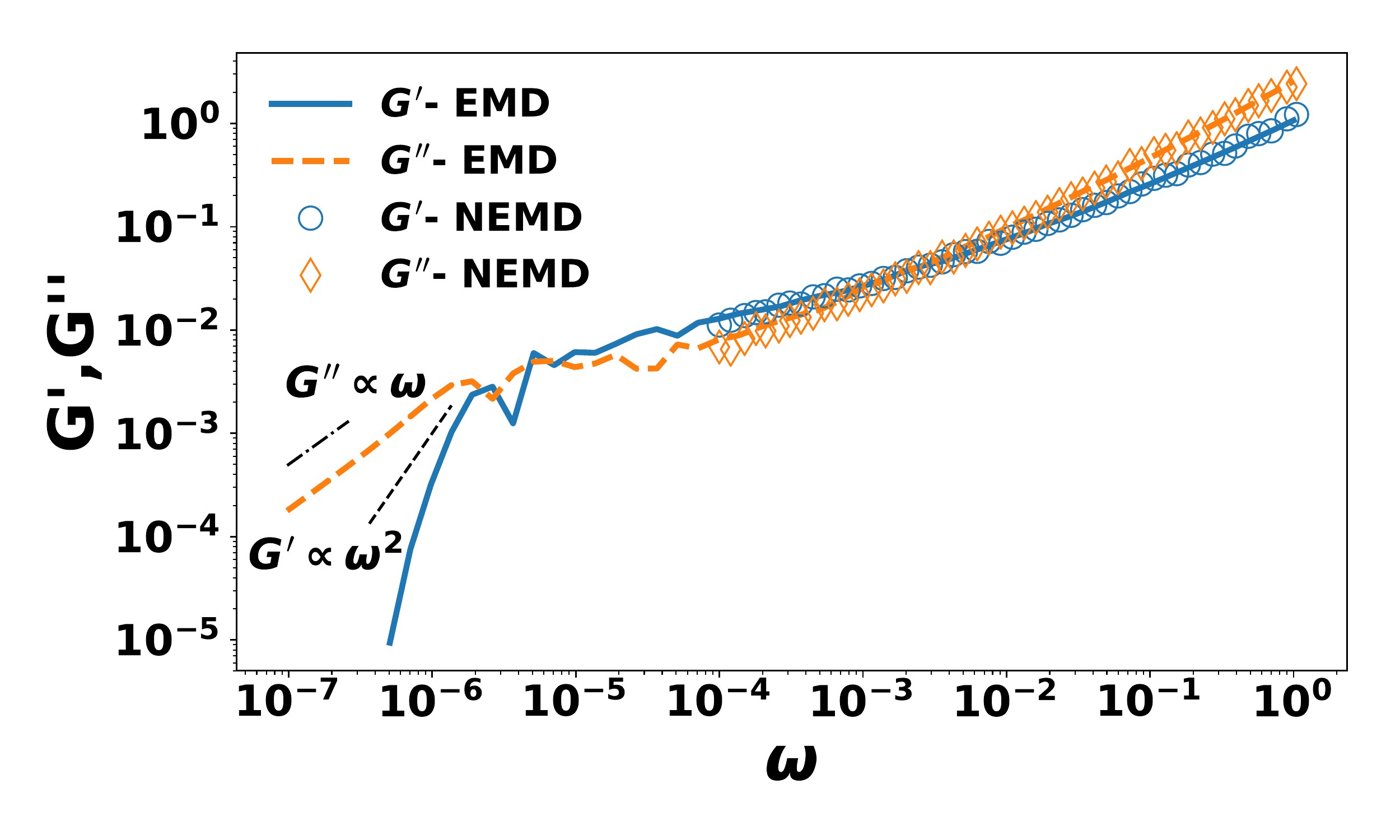}
	\caption{$G'$ and $G''$ using EMD/GK and NEMD for $N = 350$.}
	\label{fig:350_GG}
\end{figure}

\Cref{fig:350_GG} shows $G'$ and $G''$ for the longest chain species $N=350$ studied. Different from the shorter unentangled chains in \cref{fig:25_50_GG}, the entangled chains display crossovers between the $G'$ and $G''$ profiles. Two crossovers are observed in the frequency range studied. The first cross over at $\omega\sim\mathcal{O}(10^{-6})$ is at the same order of magnitude as $1/\tau_\text{d}$ -- the disentanglement time $\tau_\text{d}=1.74\times 10^{6}$ was determined from the monomer mean square displacement (MSD) curve for the same $N=350$ chains in \citet{adeyemi2021dynamics}. Crossover at $\omega\sim 1/\tau_\text{d}$ was also commonly found in experimental systems.\cite{rubinstein2003polymer}.
The second crossover, as also expected from experiments, should appear at $\omega\sim 1/\tau_\text{e}$. In our simulation, the corresponding crossover is found at $\omega\sim 2\times 10^{-3}$, whereas $\tau_\text{e}$ is $3.43 \times 10^3$ as determined, again, from MSD~\citep{adeyemi2021dynamics} -- i.e., $1/\tau_\text{e}\approx 3\times 10^{-4}$. The two values differ by a factor of 6 to 7. We note that the difference of this magnitude is not uncommon even between $\tau_\text{e}$ values measured from different experimental techniques\cite{liu2007direct}.
In addition, since $N=350$ is not long enough for the chains to be fully entangled -- as reflected by the lack of a fully developed stress plateau, quantitative discrepancies with characteristics of fully entangled polymers in experiments are excepted.\par
\citet{likhtman2007linear} also reported the first crossover between $\omega=10^{-6}$ and $10^{-5}$. Their $G'$ and $G''$ profiles appear smoother than ours in the terminal regime. This can be attributed to their use of Maxwell modes for fitting the $G(t)$ profile which inherently cannot capture the oscillations in the $G(t)$ profile -- either the short-time oscillation caused by bond fluctuations or the long-time oscillation caused by statistical uncertainty. Our fitting used piecewise linear functions (Appendix \ref{appendix_A}), which preserves all oscillations in the relaxation modulus.
We may as well obtain smooth terminal-regime profiles if we filter the $G(t)$ profile at the long-time limit before its conversion to $G'$ and $G''$.
\citet{likhtman2007linear}, however, were not able to identify the second crossover, the one corresponding to $1/\tau_\text{e}$, unless the system density is significantly raised.
Finally, we again note the excellent agreement between EMD and NEMD results in \cref{fig:350_GG}. Both methods predict the second crossover at the same position, although our NEMD did not cover sufficiently low frequency to reach the first crossover. \par
\begin{figure}
	\centering
	\includegraphics[width=3.5in, trim=0 0 0 0, clip]{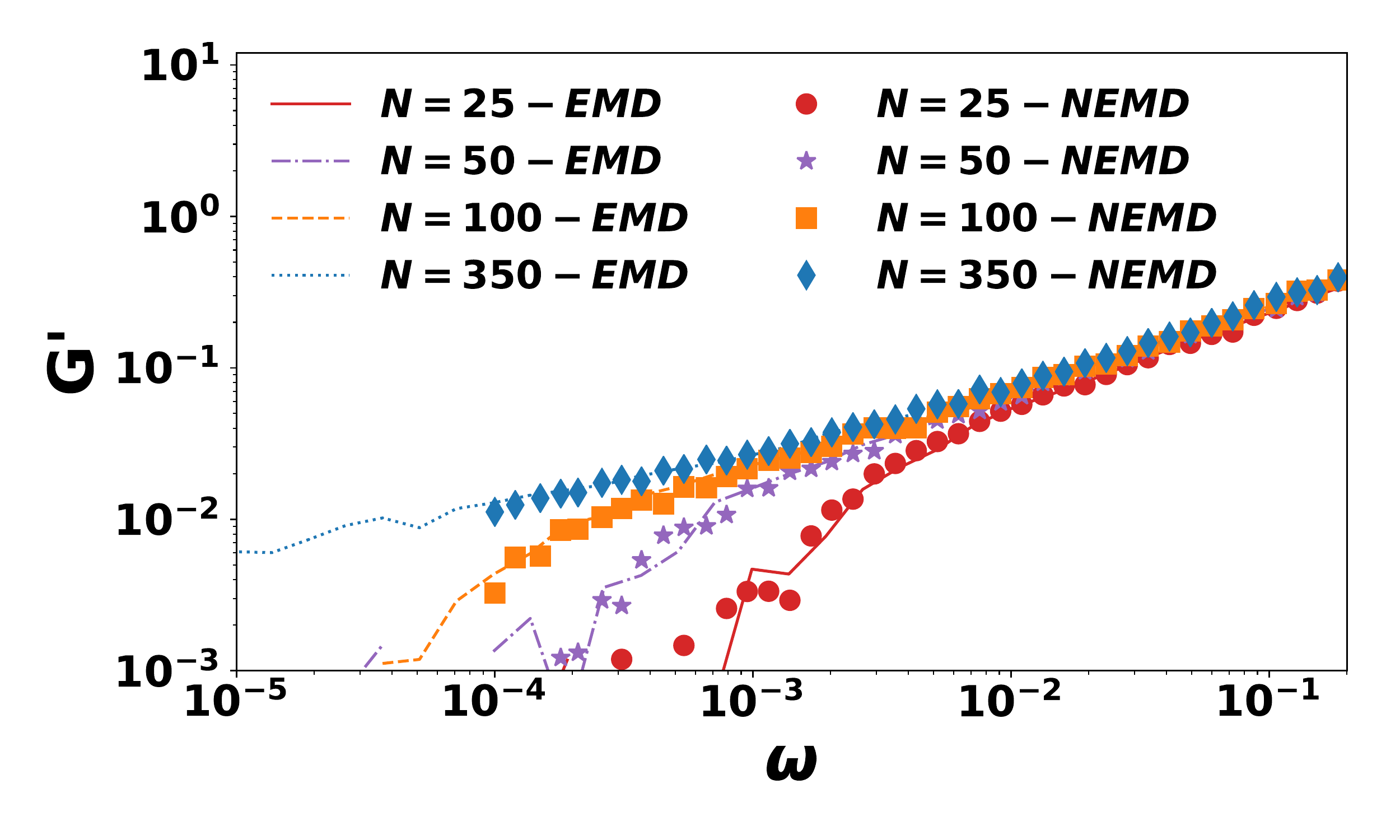}
	\caption{$G'$ for $N = 25, 50, 100$ and $350$ chains using EMD/GK and NEMD.}
	\label{Gpr_emd_nemd}
\end{figure}
\begin{figure}
	\centering
	\includegraphics[width=3.5in, trim=0 0 0 0, clip]{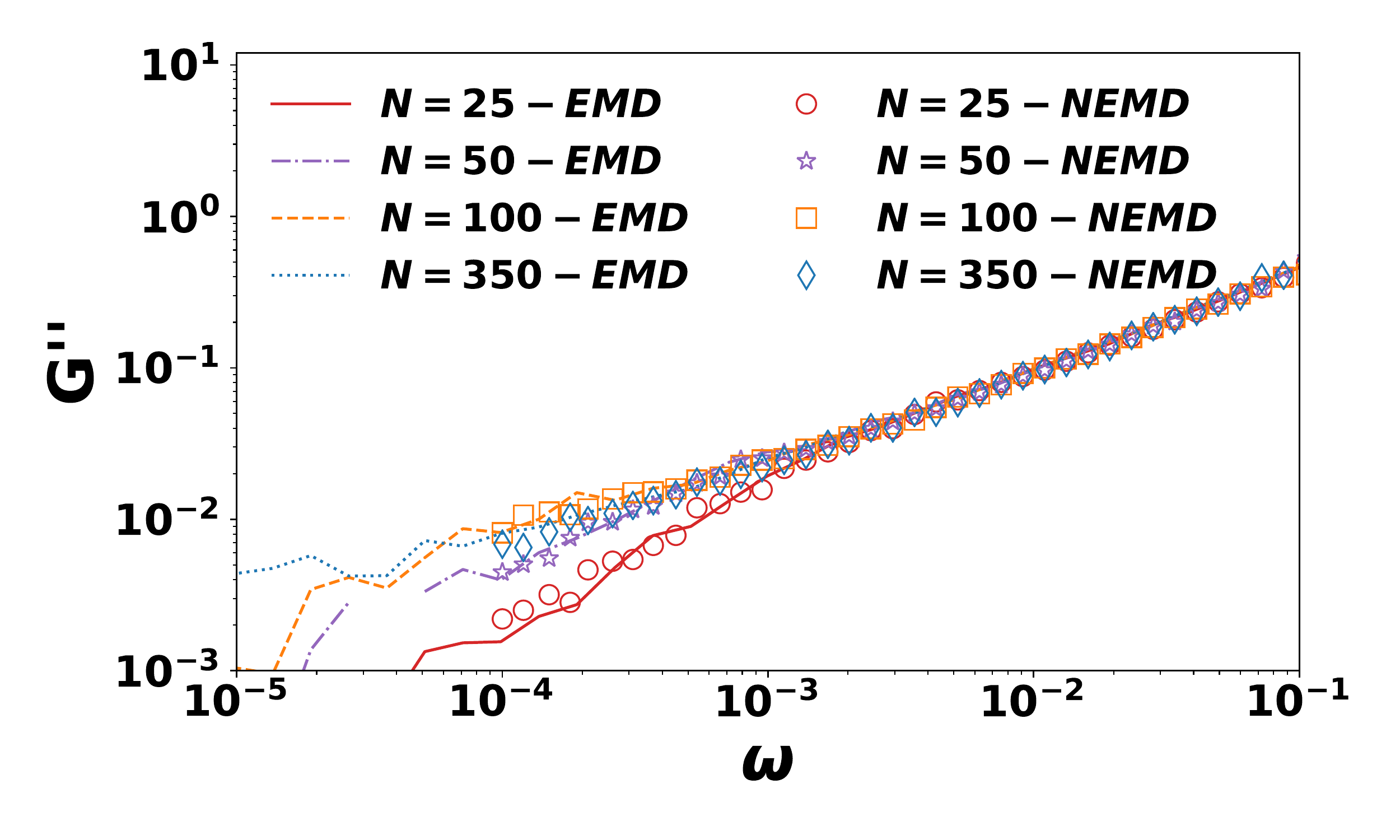}
	\caption{$G''$ for $N = 25, 50, 100$ and $350$ chains using EMD/GK and NEMD.}
	\label{Gpdr_emd_nemd}
\end{figure}

\Cref{Gpr_emd_nemd,Gpdr_emd_nemd} show the $G'$ and $G''$ results for all chain lengths. For $G'$, the curves all collapse on themselves at higher frequencies. At lower frequencies, the magnitude of $G'$ increases with increasing chain length. Entanglement effects are clearly noticeable in the $N=350$ case, where the profile decays with a lower slope at $\omega\lesssim\mathcal{O}(10^{-3})$.
It, however, falls short of developing a fully flat plateau. The slower decay allows the $G'$ profile to intersect the $G''$ profile in that frequency range (\cref{fig:350_GG}). In comparison, the unentangled species ($N=25$ and $50$) decays at faster rates as they approach the terminal regime. Signs for entanglement cannot be clearly identified from the $G''$ profiles (\cref{Gpdr_emd_nemd}).\par
In all cases, NEMD and EMD/GK results are in excellent agreement for the frequency range covered by our NEMD simulations, which provides mutual validation between these two methods. For NEMD, it is clear that, with a proper data processing procedure, one can obtain reliable results with much fewer cycles (25 in this study) than previous reports.
For EMD, its application using the GK relation has been plagued by the strong statistical noise.
\citet{likhtman2007linear} has showed that the multi-tau correlator method can effectively suppress the noise and render smooth $G(t)$ profiles. Its success, however, builds on the aggressive filtering, using extended averaging windows, at the long-time end of the TACF.
The effects of such filtering on the quantitative accuracy of the results were not known, until our direct comparison with NEMD establishes its validity.\par
With $G'$ and $G''$, we can calculate the complex modulus
\begin{gather}
	G^*\equiv\sqrt{G^{\prime 2}+G^{\prime\prime 2}}
\end{gather}
and then the complex viscosity
\begin{gather}
	\eta^*\equiv\frac{G^*}{\omega}
\end{gather}
to test the validity of the Cox-Merz rule. The steady shear viscosity $\eta$ was obtained by running the NEMD simulation of steady shear flow at different shear rates $\dot\gamma$. For each $\dot\gamma$, the first $10^{5}$ TUs of the shear stress time series was discarded and the following $1.5\times 10^{5}$ was averaged to be used in the shear viscosity calculation. Uncertainty was estimated by dividing the retained part into three blocks of equal length and the standard error of viscosity values from those blocks are reported.\par
\begin{figure}
	\centering
	\includegraphics[width=3.5in, trim=0 0 0 0, clip]{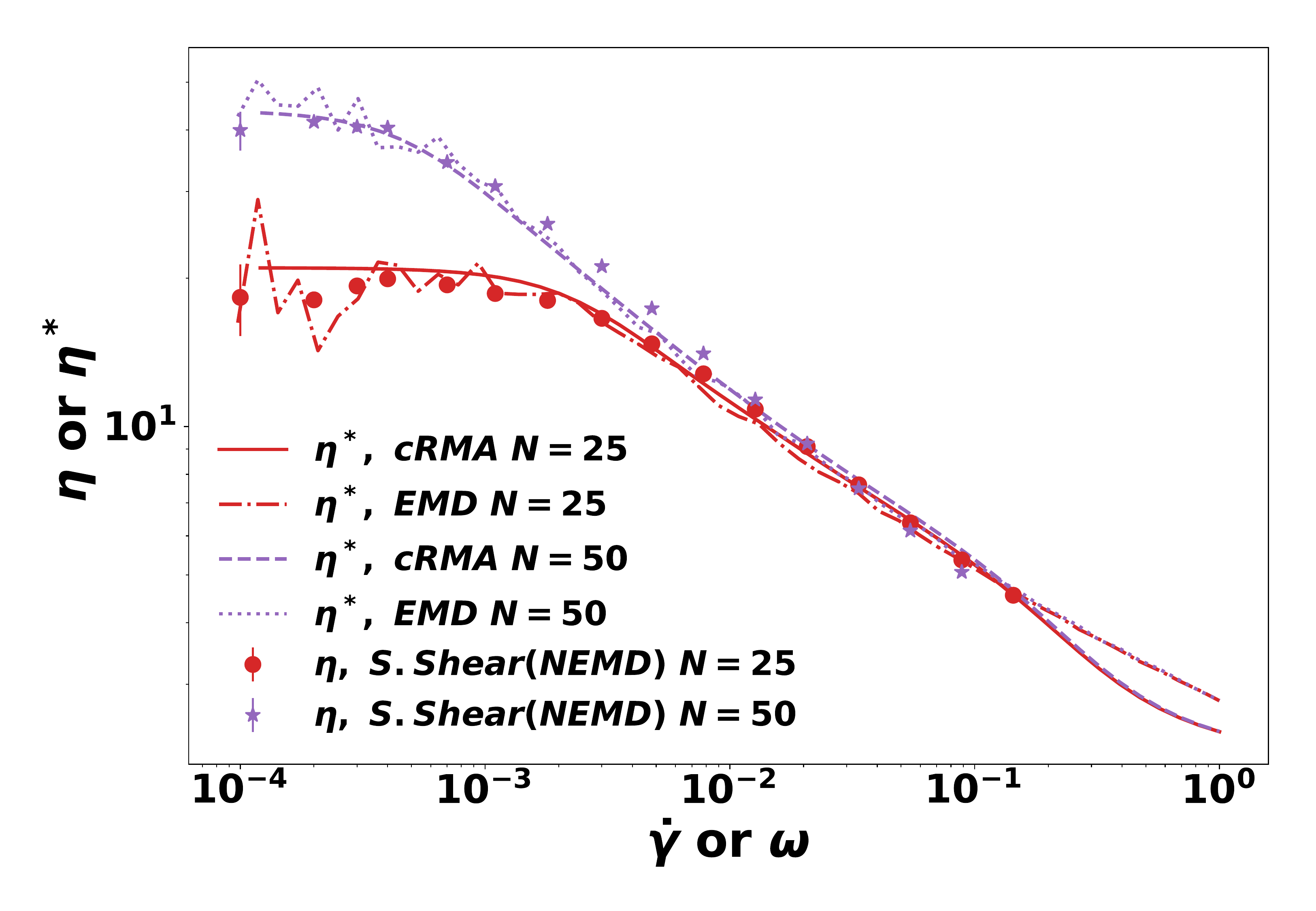}
	\caption{Comparison of the complex viscosity $\eta^*(\omega)$, from EMD/CK and cRMA approaches, with the shear viscosity $\eta(\dot\gamma)$, from NEMD of steady shear flow conditions ($N = 25$ and $N = 50$). 
	Error bars are shown for the latter but only when they are larger than the marker size.}
	\label{fig:dynamic_viscosity}
\end{figure}
\Cref{fig:dynamic_viscosity} plots the steady shear viscosity $\eta(\dot\gamma)$ in comparison with $\eta^*(\omega)$ for the unentangled chain species. Only EMD/GK and cRMA results are plotted for $\eta^*$. The NEMD/SAOS results are very close to EMD/GK (as reflected in their numerically equivalent $G'$ and $G''$ results) and thus omitted for clarity. The viscosity profiles show typical behaviors of polymer melts, including a Newtonian plateau at the low shear end and shear-thinning at higher shear rates. It is clear that $\eta(\dot\gamma)$ stays close to $\eta^*(\omega)$ for the entire range tested, indicating the general applicability of the Cox-Merz rule to the KG model chains. Both EMD/GK and NEMD/steady shear are subject to larger statistical uncertainty at the low $\omega$ or $\dot\gamma$ end, while the cRMA approach gives smooth and accurate results for unentangled chains.
\begin{figure*}
	\centering
	\begin{subfigure}[b]{0.48\textwidth}
		\centering
		\includegraphics[width=\textwidth]{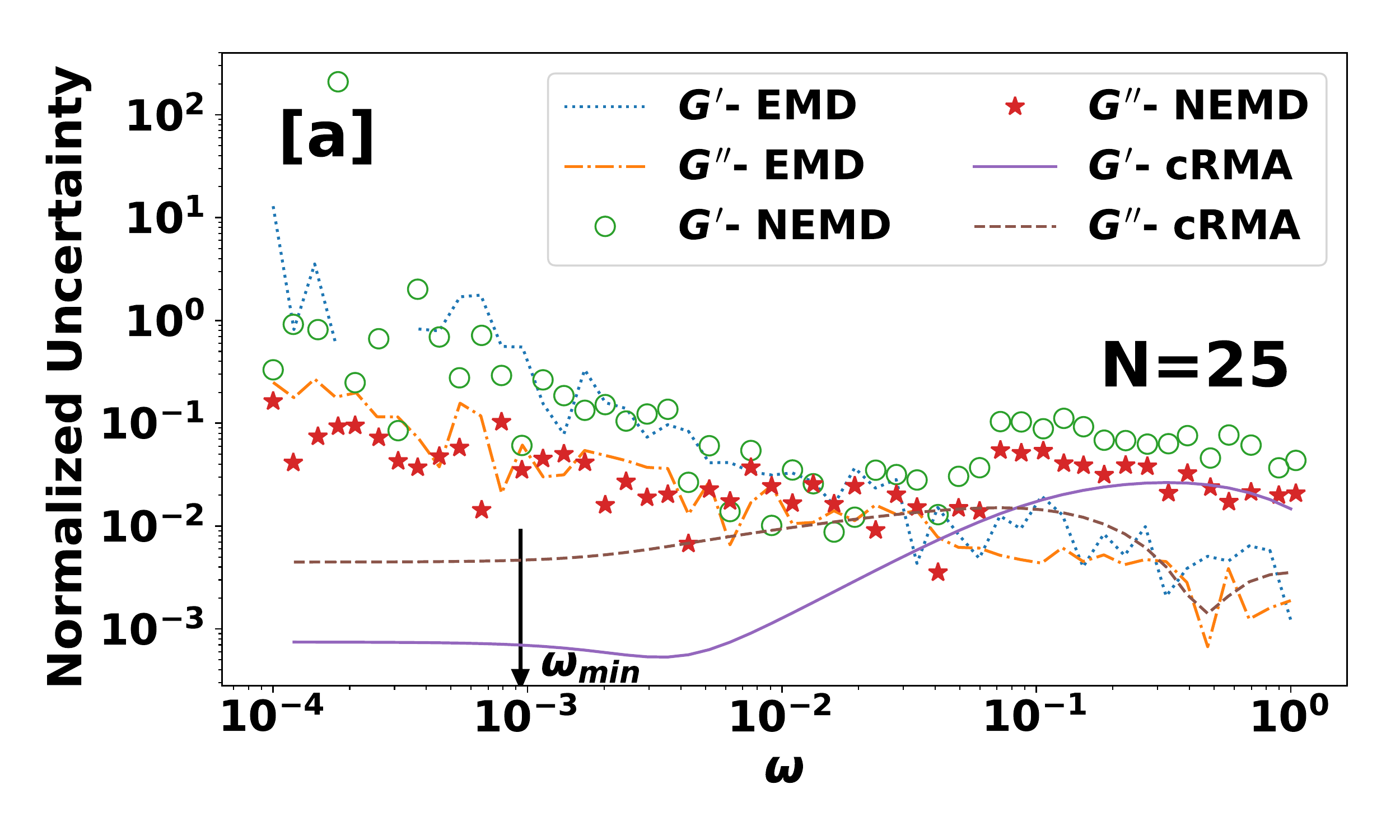}
		\phantomsubcaption{}
		\label{fig:25_unc}
	\end{subfigure}
	\begin{subfigure}[b]{0.48\textwidth}
		\centering
		\includegraphics[width=\textwidth]{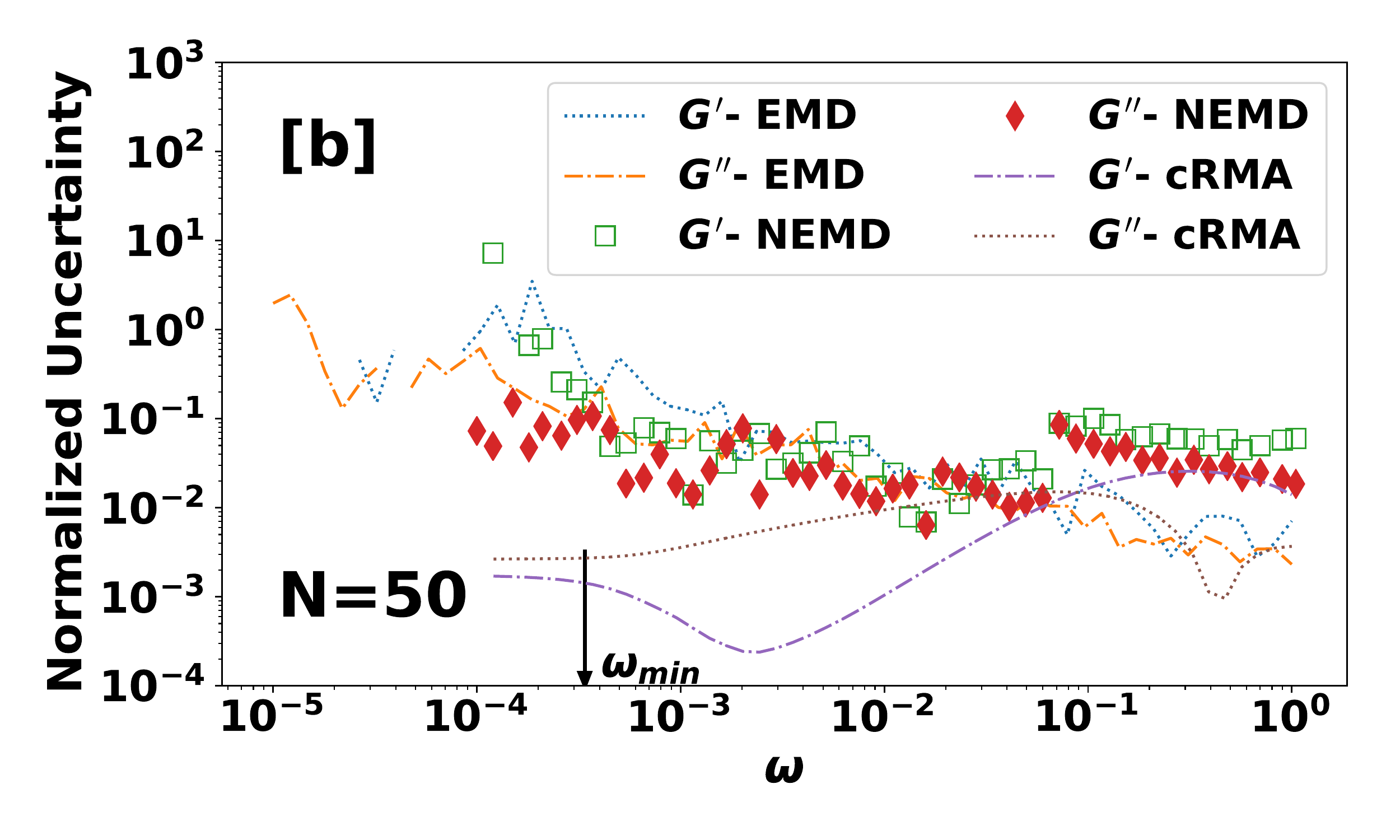}
		\phantomsubcaption{}
		\label{fig:50_unc}
	\end{subfigure}
	\begin{subfigure}[b]{0.48\textwidth}
		\centering
		\includegraphics[width=\textwidth]{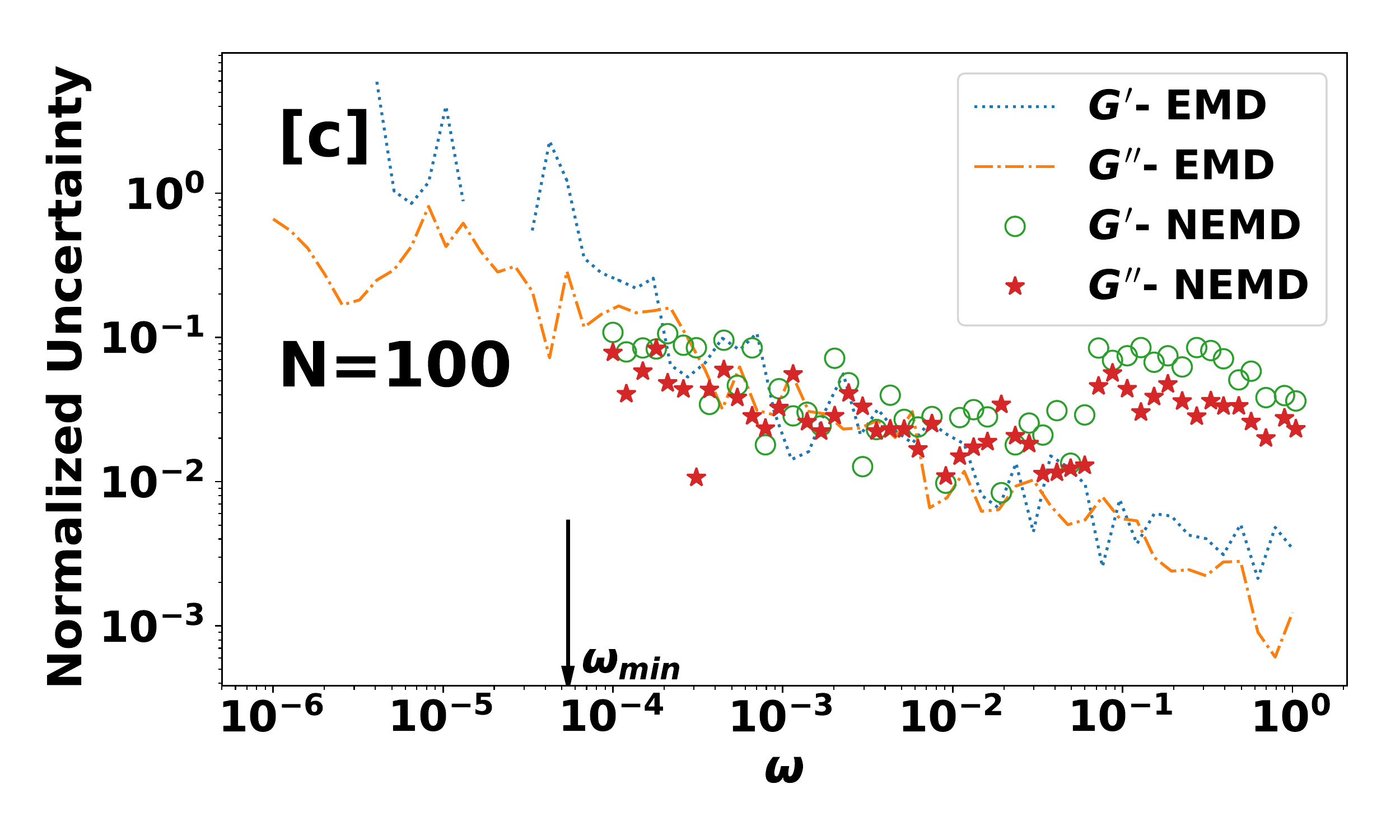}
		\phantomsubcaption{}
		\label{fig:100_unc}
	\end{subfigure}
	\begin{subfigure}[b]{0.48\textwidth}
		\centering
		\includegraphics[width=\textwidth]{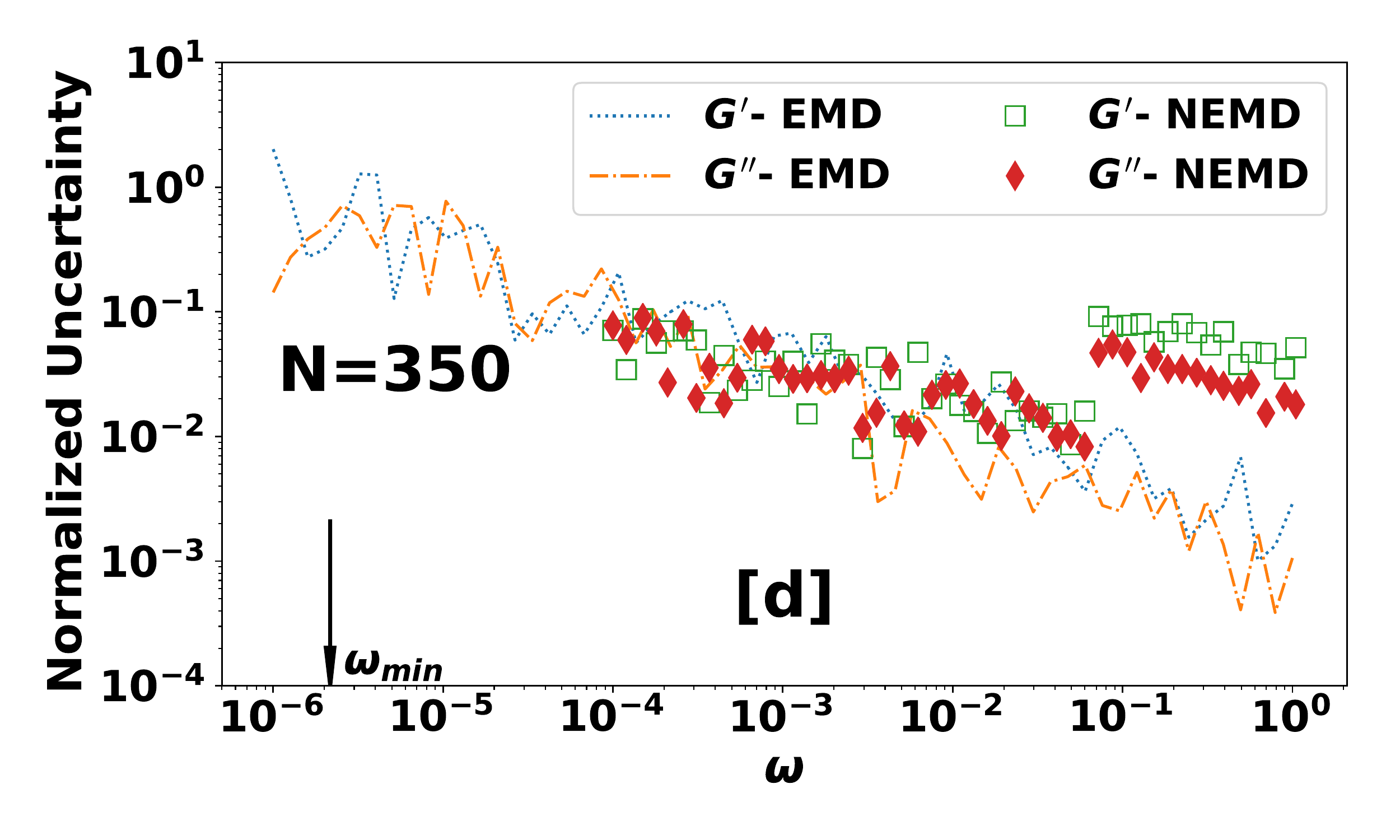}
		\phantomsubcaption{}
		\label{fig:350_unc}
	\end{subfigure}
	\caption{Uncertainty of $G'$ and $G''$ from EMD/GK, NEMD, and (for unentangled chains only) cRMA normalized by the estimated $G'$ and $G''$ values: (a) $N = 25$, (b) $N=50$, (c)$N=100$, and (d) $N=350$.}
	\label{fig:uncertainty_analysis}
\end{figure*}
\subsection{Discussion: Accuracy and Cost}\label{sec:cost_disc}
Results presented so far have established that, with proper noise reduction and data processing procedures, both EMD and NEMD give quantitatively reliable results for $G'$ and $G''$. The question now becomes which method should one choose for obtaining the most accurate results with limited computational resources. \par
\Cref{fig:uncertainty_analysis} shows the statistical uncertainty in the $G'$ and $G''$ values calculated from all three methods using our standard simulation lengths reported in Sec. \ref{sec:methods}. For EMD/GK and cRMA, uncertainty is straightforwardly estimated from the standard error of results from independent trajectories. As shown in \cref{tab:timescales}, three to five independent EMD runs were performed for each case. For NEMD/SAOS, the 25-cycle time series used for each frequency was divided into five equal blocks  (with five cycles in each). Each block of time series undergoes the DFT analysis to obtain its own $G'$ and $G''$ values and the uncertainty is reported as the standard error between single-block results. The reported uncertainty magnitudes in \cref{fig:uncertainty_analysis} are all normalized by the corresponding $G'$ or $G''$ values -- i.e., they are reported as relative errors. \par
Accuracy of $G'$ and $G''$ results must be discussed in the frequency range of relevance, which varies with chain length. We define the maximum stress relaxation time $\tau_\text{max}$ as the time for $G(t)$ to first drop to $10^{-3}$ (see \cref{fig:Gt_plot}) and listed the timescale in \cref{tab:timescales} for different chain lengths. We note that for $N=350$, $\tau_\text{max}$ is much longer than its Rouse time $\tau_\text{R}=1.66\times 10^5$ (as determined from MSD~\cite{adeyemi2021dynamics}) due to entanglement effects, whereas for $N=25$ and $50$, $\tau_\text{max}$ is very close to their respective $\tau_R$ (which can be estimated from the $\tau_\text{R}$ of the $N=350$ case using $\tau_\text{R}\propto N^2$). The standard EMD simulation length chosen for each independent run is one to two orders of magnitude longer than $\tau_\text{max}$ to ensure that the stress TACF has multiple independent segments to average over for the longest time scale of interest. We then mark $\omega_\text{min}\equiv 1/\tau_\text{max}$ as the minimum frequency of interest for each chain length in \cref{fig:uncertainty_analysis}. \par
For unentangled cases ($N=25$ and $50$), cRMA is clearly more accurate than both other methods, especially at the low-frequency end, where both EMD/GK and NEMD suffer from strong fluctuations, the statistical error from cRMA is well below $1\%$. \par
Between EMD/GK and NEMD, there is notable difference in the frequency dependence of uncertainty. The GK relation relies on the stress TACF to calculate $G(t)$. For EMD simulation of a given duration, there are more shorter independent segments to average over than longer ones. As a result, at $\omega\gtrsim 10^{-2}$, its error is rather low -- no more than a few percent, while each EMD case see its largest error at the low frequency end. Uncertainty from NEMD is less dependent on frequency and fluctuates more or less in the $10^{-2}$ to $10^{-1}$ range. In \cref{fig:25_unc} and \cref{fig:50_unc}, the error does seem to grow above $10\%$ at the low frequency end, but that is likely due to the frequency dropping below $\omega_\text{min}$, where the complex modulus magnitudes are vanishingly small and no longer of significant interest. It appears that for NEMD, the uncertainty depends mostly on the number of cycles included in the statistics which was set to be the same at different frequencies. \par 
To compare the efficiency between EMD and NEMD, we first look at the $N=350$ case (\cref{fig:350_unc}), where the simulation cost, measured in terms of the total number of MD time steps used in the statistics (all three independent runs for EMD and 25 cycles at all frequencies for NEMD), is controlled to be nearly the same. From \cref{fig:350_unc}, the statistical errors from both methods are comparable in a wide frequency range of $10^{-4}\lesssim\omega\lesssim 10^{-1}$. The advantage of EMD is clear at $\omega \gtrsim 10^{-1}$, where its error drops below $1\%$, but NEMD remains acceptable at below $10\%$. The higher error from NEMD at high frequency is attributed to the declining effectiveness of the pre-averaging step applied to the stress signal. To avoid contamination of stress signal at the imposed frequency, we set the pre-averaging block size to $1/100$ of the oscillation period. As the imposed frequency increases, the block size diminishes and becomes less effective at noise removal.
One may easily improve the accuracy at high frequencies by running more cycles, which would not introduce substantial extra cost due to the shorter periods there. 
Per \cref{fig:gg_uncertainty}, increasing to 100 cycles is estimated to reduce the error in $G'$ by half. (Although \cref{fig:gg_uncertainty} used a block size of 1 cycle for error estimation -- versus 5 cycles used in \cref{fig:uncertainty_analysis}, we have confirmed that the dependence of error on $N_\text{cycle}$ is not sensitive to the block size.) \par
Limitation of NEMD is more obvious at the low frequency end. The frequency range swept by NEMD in this study goes down to $10^{-4}$, which leaves nearly two decades of lower frequencies that are still of interest (i.e., $>\omega_\text{min}$) uncovered.
By contrast, the same set of EMD data can be used to generate $G'$ and $G''$ of any frequency without additional computational cost. Of course, for limited EMD simulation length, statistical uncertainty increases with decreasing frequency, but as far as results in \cref{fig:350_unc} are concerned, the error remains at $\sim 10\%$ for most of the $\omega\sim\mathcal{O}(10^{-5})$ decade. To capture the same decade using NEMD, the computational cost would be 10 times as high as that of the $\mathcal{O}(10^{-4})$ decade -- i.e., the overall NEMD simulation cost must increase by an order of magnitude. Based on \cref{fig:gg_uncertainty}, one may propose to accept slightly higher uncertainty and run the lowest frequencies with fewer cycles, which nonetheless would still require significantly higher computational cost.\par
The conclusion is similar at $N=100$ (\cref{fig:100_unc}), where the lowest frequency swept by NEMD is closer to $\omega_\text{min}$. NEMD also shows similar uncertainty level as EMD except at $\omega\gtrsim 10^{-1}$ where the advantage of EMD is clear. Note that this equivalence in performance between these two methods is built on substantially higher computational cost in NEMD. Recall that the total computational cost of NEMD in this study does not change with chain length. For $N=100$, the cost of EMD (\cref{tab:timescales}) is only one third that of NEMD. This, however, does not mean that EMD is three times better -- everything else the same, increasing the data size by a factor of three would lead to a factor of $\sqrt{3}$ reduction in the uncertainty, which is not big compared with fluctuations between data points in \cref{fig:100_unc}. The advantage of EMD is smaller for shorter chains ($N=25$ and $50$ in \cref{fig:25_unc} and \cref{fig:50_unc}). In both cases, NEMD offers similar statistical accuracy as EMD except, again, at the high-frequency end. The total cost of NEMD is higher by nearly one order of magnitude, but part of the low frequency data fall below $\omega_\text{min}$. If we only count NEMD runs at $\omega\geq\omega_\text{min}$, the total computational cost would be comparable to EMD at $N=25$.\par
\begin{figure}
	\centering
	\begin{subfigure}[b]{0.48\textwidth}
		\centering
		\includegraphics[width=\textwidth]{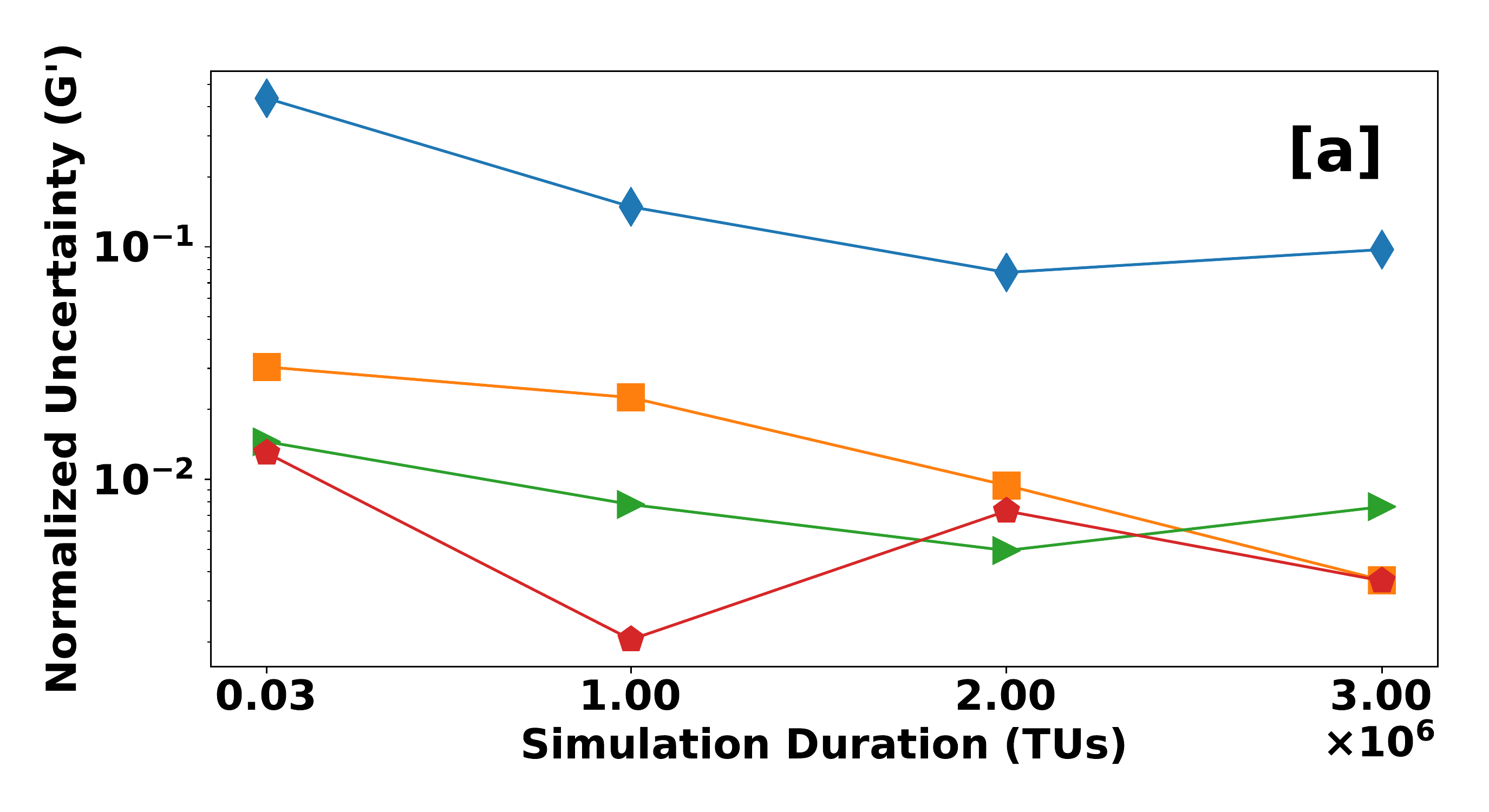}	
		\phantomsubcaption
		\label{fig:emd_dependence_gpr}	
	\end{subfigure}
	\begin{subfigure}[b]{0.48\textwidth}
		\centering
		\includegraphics[width=\textwidth]{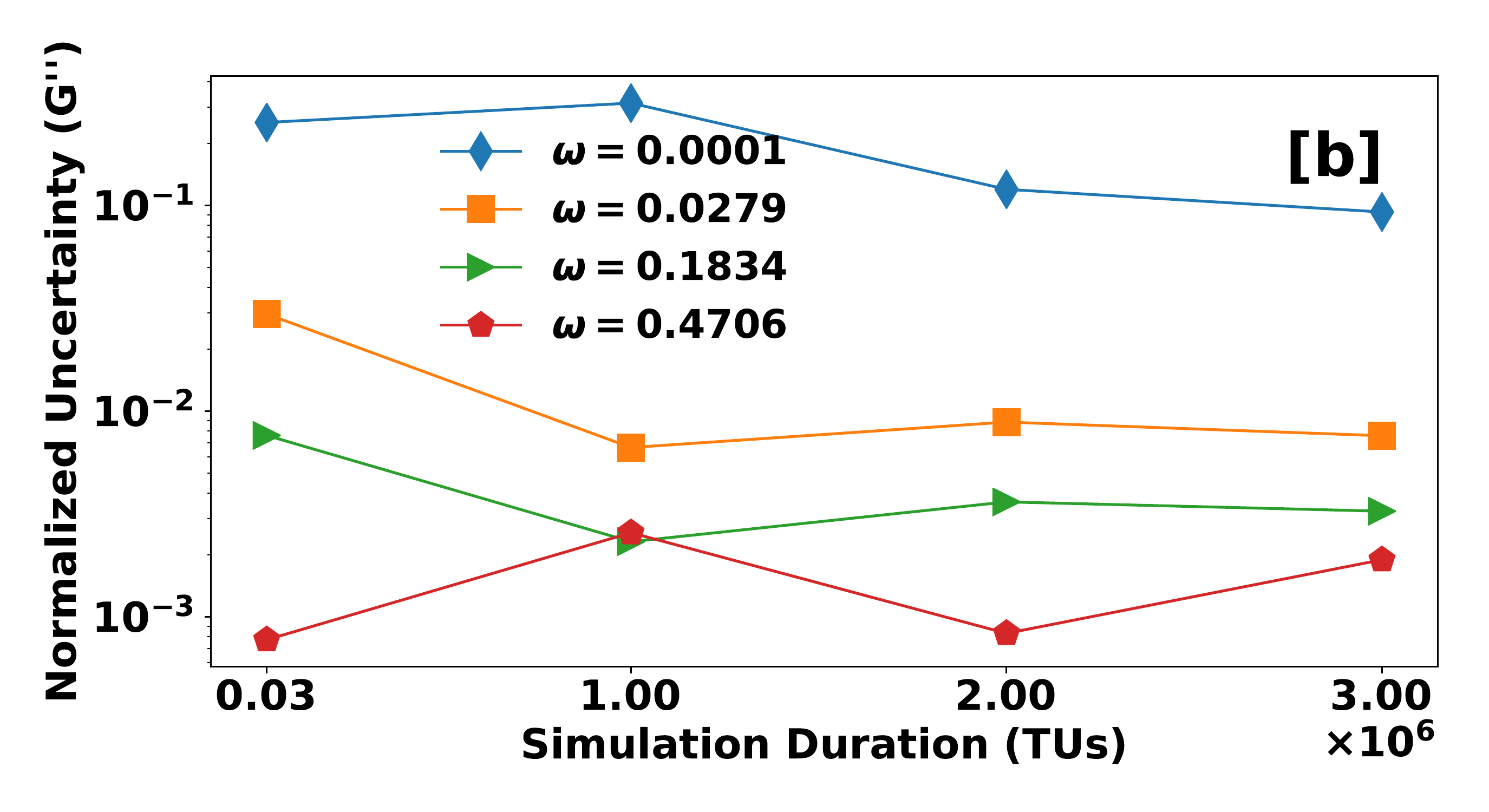}
		\phantomsubcaption 
		\label{fig:emd_dependence_gdpr}
	\end{subfigure}
	\caption{Dependence of normalized uncertainty in EMD/GK results on the duration of each independent simulation run (out of three used in the statistics of $N=350$): (a) $G'$ and (b) $G''$. The longest run shown in the figure  (with $3\times 10^6$ TUs) is the standard duration used in the study.}	
	\label{fig:emd_dependence}
\end{figure}

Our analysis shows that, contrary to many's belief, EMD using the GK relation and multi-tau correlator method not only provides accurate results for linear viscoelastic properties, it also appears to be more efficient in some cases, especially for longer chains where the need of covering lower frequencies puts higher burden on NEMD.
For EMD, in theory, meaningful results at all frequencies can be generated with a single run that covers the longest relaxation time. In practice, EMD is equally constrained by the limited simulation duration in the long-time (low-frequency) end of the spectrum. \Cref{fig:emd_dependence} shows the variation of the normalized statistical error of EMD if we shorten the duration of each independent run to $1/10$, $1/3$, and $2/3$ of the standard duration (\cref{tab:timescales}). It is clear that as the simulation gets shorter, accuracy at lower frequencies is first affected. For example, with a 10-fold increase in simulation length, the error in $G'$ reduces by a factor of 3 to 5 (\cref{fig:emd_dependence_gpr}), which is comparable to the factor of $\sqrt{10}$ expected.\par

The advantage of EMD is that information on different frequencies is contained in the same time series, whereas NEMD would require a new simulation even for a slightly different frequency.
Although EMD seems more susceptible to statistical noise, which is easier to remove in NEMD because the frequency of the primary signal is known \textit{a priori}, this weakness is partially lessened by the success of the multi-tau correlator method.
The net outcome is thus \RevisedText{an} advantage in favor of EMD when computing the complete spectrum of linear viscoelasticity is the goal.
The real advantage of NEMD lies in its flexibility. For example, one may easily save half of the computational cost by dropping every other frequency level covered. It would also be preferred when only a certain frequency range is of interest or lower accuracy is permissible at certain frequencies. The latter is because it allows the user to independently adjust the accuracy at different levels by changing the number of cycles used. 

\RevisedText{
The comparison between these two approaches is determined by the balance of cost between prolonging EMD simulation for reduced statistical uncertainty and repeating NEMD simulation at different frequency levels.
This balance may shift for a different system or a different model.
We have already observed that the advantage of EMD vanishes as the chain length decreases. For non-polymeric simple liquids, NEMD may as well be the more efficient approach given the much shorter frequency range that needs to be covered.
In this study, we have only tested the KG model. Chemically realistic atomistic molecular models are likely to produce much stronger stress fluctuations, posing extra challenges for noise reduction in both approaches. In particular, whether the multi-tau correlator method can still sufficiently reduce the noise in EMD to keep its relative advantage remains to be tested.
Over the past two decades, there has been a growing trend of developing coarse-grained molecular models that map reversibly to atomistic models~\citep{xi2019molecular,Peter2009}. Such models often map one or more polymer repeating units into a single super atom and the effective interactions between such super atoms are generally softer than those in both atomistic models and the KG model. It is thus possible that such models are less susceptible to stress fluctuations. 
Meanwhile, coarse-graining is known to cause the artificial acceleration of the system dynamics in MD, which obviously alters the calculated viscoelastic properties. The problem can be countered by explicitly introducing friction drag and random forces to the model. The effects of this treatment on stress fluctuations, which remain unknown, introduce another variable in the balance of cost between EMD and NEMD.
The reader is referred to \citet{xi2019molecular} for detailed discussion on the application of coarse-graining in the molecular simulation of polymer rheological properties. 
}

\section{Conclusions}
In this study, we compared equilibrium and non-equilibrium MD approaches for computing the linear viscoelastic properties of polymer melts, using a KG bead-spring chain model with chain lengths that range from the unentangled ($N = 25$ and $50$) to the marginally and moderately entangled ($N = 100$ and $N=350$) regimes. For EMD, the primary focus was on the Green-Kubo (GK) approach, but, for unentangled chains, we also tested a corrected Rouse mode analysis approach in which short-time GK results were introduced to supplement the stress relaxation modulus calculation from the Rouse model. We showed that with proper data processing and noise reduction procedures, all these approaches produced quantitatively equivalent results for $G'$ and $G''$. For EMD with the GK relation, the multi-tau correlator method effectively removes the noise while preserving the quantitatively accurate relaxation dynamics. Numerical integration of the Fourier integrals with the relaxation modulus $G(t)$ approximated by a piecewise linear function faithfully converts the results to complex moduli. For NEMD, we applied DFT to extract $G'$ and $G''$ from the pre-averaged stress signal and showed that 25 cycles at each frequency is sufficient to obtain statistically meaningful results. The simulation length is much shorter than previously reported in the literature which significantly reduces the computational expense needed to obtain a representative spectrum. In addition, we found that the strain amplitude of the imposed oscillatory shear must be carefully chosen for different frequency levels to avoid non-linear effects. \par
Comparing the statistical uncertainty of these methods, we found that, despite the common perception that the EMD/GK approach is more strongly influenced by stress fluctuations, it offers at least equally accurate and, sometimes, more accurate results than NEMD when the same total simulation time is used. The advantage of NEMD is its flexibility especially when only a limited frequency range is of interest.
The cRMA method relies on the accuracy of the Rouse model but, at least for the KG model in the unentangled regime, it offers highly accurate results.

\appendix
\section{Numerical Evaluation of the Fourier Integral}\label{appendix_A}
Assume we are given $G(t)$ values at a series of discrete points: $G_1, G_2, ..., \text{and } G_k$, where $G_k$ represents the value of $G(t)$ at the $k$-th temporal grid point $t_k$. The data points do not have to be equally spaced apart. Indeed, in this study, the discretized $G(t)$ points came from the multi-tau correlator method, which by construction uses a non-uniform temporal grid and its spacing increases with $t$. We used the multi-tau correlator output series of $G_k$ without modification. \par
Note that evaluating \cref{eq:Gprime,eq:Gdprime} is equivalent to performing the Fourier integral
\begin{equation}\label{eq:compact_FI}
	I \equiv \int_0^\infty G(t)\exp(-i\omega t) dt.
\end{equation}
For its numerical evaluation, we follow the method in \citet{luyben1973process} and divide the integral into sub-integrals of individual grid intervals -- i.e., $\Delta t_k \equiv t_k - t_{k-1}$. \Cref{eq:compact_FI} is then written as the summation of sub-integrals $I_k$:
\begin{equation}\label{eq:FIT_approx}
	I = \sum_{k=1}^{N}\left(\int_{t_{k-1}}^{t_k} G(t) \exp(-i \omega t)dt \right) \equiv \sum_{k=1}^N I_k 
\end{equation}
We now approximate $G(t)$ in each interval $t_{k-1}$ to $t_k$ with a linear function (higher order polynomials can be used to improve the accuracy):
\begin{align}\label{eq:str_line}
 G(t) 	& \approx \phi_k(t)  \notag \\
	&  = \alpha_{0k} + \alpha_{1k}(t-t_{k-1}) \qquad \text{for} \quad t_{k-1} < t < t_k 
\end{align}
where $\alpha_{1k}$ is the slope of the line over the $k$-th interval
\begin{equation} \label{eq:alpha_1k}
	\alpha_{1k} = \frac{G_k - G_{k-1}}{\Delta t_k}
\end{equation}
and $\alpha_{0k}$ is the value of $\phi_k$ at the beginning of the interval
\begin{equation}\label{eq:alpha_0k}
	\alpha_{0k} = G_{k-1}.
\end{equation}
The constants $\alpha_{0k}$ and $\alpha_{1k}$ change with each interval. Inserting \cref{eq:str_line} into \cref{eq:FIT_approx} gives
\begin{align}\label{eq:Ik_approx}
	& I_k \approx \int_{t_{k-1}}^{t_k}[\alpha_{0k}+\alpha_{1k}(t-t_{k-1})]\exp(-i \omega t)dt 
\end{align}
which can be evaluated analytically. Integrating \cref{eq:Ik_approx} by parts and substituting $\alpha_{0k}$ and $\alpha_{1k}$ by \cref{eq:alpha_1k} and \cref{eq:alpha_0k} give
\begin{gather}
\begin{split}
	I_k &{}\approx \frac{G_{k-1}}{i\omega}\left(\exp(-i \omega t_{k-1})-\exp(-i \omega t_k)\right) \\
	&\qquad -\frac{G_k - G_{k-1}}{\Delta t_k}\frac{\Delta t_k}{i \omega}\exp(-i \omega t_k) \\
	&\qquad + \frac{G_k - G_{k-1}}{\Delta t_k \omega^2}\left(\exp(-i \omega t_k)-\exp(-i \omega t_{k-1})\right)\\
	& = G_k\left(-\frac{\exp(-i \omega t_k)}{i \omega}+\frac{\exp(-i \omega t_k)-\exp(-i \omega t_{k-1})}{\omega^2 \Delta t_k}\right)\\
	&\qquad +G_{k-1}\left(\frac{\exp(-i \omega t_{k-1})}{i \omega}-\frac{\exp(-i \omega t_k)-\exp(-i \omega t_{k-1})}{\omega^2 \Delta t_k}\right).
\end{split}
\end{gather}
Extracting $\exp(-i\omega t_{k-1})$ and noting that $\Delta t_k = t_k-t_{k-1}$, we obtain
\begin{gather}
\begin{split}
	I_k \approx & \exp(-i \omega t_{k-1})\biggl\{G_k \left(\frac{\exp(-i \omega \Delta t_k) -1}{\omega^2 \Delta t_k}-\frac{\exp(-i \omega \Delta t_k)}{i \omega}\right) \\
	&{}-G_{k-1}\left(\frac{\exp(-i \omega \Delta t_k) - 1}{\omega^2 \Delta t_k}-\frac{1}{i \omega}\right) \biggr\}.
\end{split}
\end{gather}
Finally, the full integral is given as
\begin{gather}
\begin{split}
	&{}\int_{0}^{\infty} G(t) \exp(-i \omega t)dt \approx \\
	&\sum_{k=1}^{N}
		\exp(-i \omega t_{k-1})\biggl\{G_k \left(\frac{\exp(-i \omega \Delta t_k) -1}{\omega^2 \Delta t_k}-\frac{\exp(-i \omega \Delta t_k)}{i \omega}\right)  \\ 
	&\qquad -G_{k-1}\left(\frac{\exp(-i \omega \Delta t_k) - 1}{\omega^2 \Delta t_k}-\frac{1}{i \omega}\right) \biggr\}. 
\end{split}
\end{gather}
\section{Data Processing for the Stress Output from Small Amplitude Oscillatory Shear (SAOS) in NEMD}\label{appendix_B}
The SAOS output (\cref{eq:sig2}) can be rewritten as
\begin{equation}
	s(t) = \frac{\sigma(t)}{\gamma_0}=G'(\omega)\text{sin}(\omega t)+G''(\omega)\text{cos}(\omega t).
\end{equation}
Assume that the total NEMD run covers $N_\text{cycle}$ \emph{whole} oscillatory cycles with a combined temporal duration of $T_\text{run}$, and $s(t)$ is stored on $N_t$ grid points with equal spacing $\Delta t$. The time mark at each grid point is 
\begin{align}
	t_j = j \Delta t = \frac{jT_\text{run}}{N_t}
\end{align}
and 
\begin{gather}
\begin{split}
	s_j \equiv s(t_j) = & G'\text{sin} \left (\frac{\omega jT_\text{run}}{N_\text{t}}\right) +G''\text{cos}\left(\frac{\omega jT_\text{run}}{N_\text{t}}\right) \\
	&(j = 0,1,...,N_t-1).
\end{split}
\end{gather}
(Note: the $j=N_t$ point is not included because we assign $s_{N_t}=s_0$ to enforce the periodicity of the time series.) The discrete Fourier transform (DFT) of the series is
\begin{gather}
\label{eq:s_k}
\begin{split}
	\hat{s}_k &= \frac{1}{N_t}\sum_{j=0}^{N_t-1} s_j \exp(-\frac{2\pi ikj}{N_t})\\
	&=\frac{1}{N_t}\sum_{j=0}^{N_t-1}\left(G'\text{sin} \left (\frac{\omega jT_\text{run}}{N_t}\right) +G''\cos \left(\frac{\omega jT_\text{run}}{N_\text{t}}\right) \right)\\
	&\qquad\left(\cos\left(-\frac{2\pi kj}{N_t}\right) + i\sin\left(-\frac{2\pi kj}{N_t}\right)\right)\\
	&=\frac{1}{N_t}\sum_{j=0}^{N_t-1}\left(G'\sin \left (\frac{2\pi k_\omega j}{N_t}\right) +G''\cos\left(\frac{2\pi k_\omega j}{N_t}\right) \right)\\
	&\qquad\left(\text{cos}\left(-\frac{2\pi kj}{N_t}\right) + i \text{sin}\left(-\frac{2\pi kj}{N_t}\right)\right)\\
	&=\frac{1}{N_t}\Biggl(\sum_{j=0}^{N_t-1}G' \sin \left(\frac{2 \pi k_\omega j}{N_t} \right)\cos\left(-\frac{2\pi kj}{N_t}\right)\\
	&\qquad+\sum_{j=0}^{N_t-1}G''\cos\left(\frac{2\pi k_\omega j}{N_t}\right) \cos\left(-\frac{2\pi kj}{N_t} \right)\\
	&\qquad+ i \Biggl(\sum_{j=0}^{N_t-1} G'\text{sin}\left(\frac{2\pi k_\omega j}{N_t}\right) \text{sin}\left(-\frac{2\pi kj}{N_t}\right)\\
	&\qquad+ \sum_{j=0}^{N_t-1}G''\cos\left(\frac{2\pi k_\omega j}{N_t}\right) \sin\left(-\frac{2\pi kj}{N_t}\right)\Biggr)\Biggr)
\end{split}
\end{gather}
%
where
\begin{equation}\label{kw}
	k_\omega \equiv \frac{\omega T_\text{run}}{2\pi}=\frac{T_\text{run}}{T_\text{cycle}}=N_\text{cycle}
\end{equation}
i.e., the total number of oscillatory cycles in the run (note that $\omega/2\pi$ equals the frequency of oscillation -- i.e., the reciprocal of the cycle period $T_\text{cycle}$). Due to the orthogonality of sine and cosine functions, for the typical situation of $0<k_\omega << N_t$, \cref{eq:s_k} is non-zero only for $k=k_\omega$ and $k=N_t-k_\omega$. The latter is equivalent to $k=-k_\omega$ due to the $2\pi$-periodicity of these functions. The non-zero modes are complex conjugates
\begin{gather}
	\hat{s}_{\pm k_\omega} =\frac{1}{2}\left(G''\mp iG'\right)
\end{gather}
containing $G'$ and $G''$ in their imaginary and real parts, respectively.
\begin{acknowledgments}
	The authors acknowledge the financial support from the Natural Sciences and Engineering Research Council of Canada (NSERC) Discovery Grants program (No.~RGPIN-2014-04903 and No.~RGPIN-2020-06774) and the allocation of computing resources awarded by Compute/Calcul Canada.
	S.Z. thanks the Canada Research Chairs (CRC) program (No.~950-229035).
	This work is made possible by the facilities of the Shared Hierarchical Academic Research Computing Network (SHARCNET: www.sharcnet.ca).
\end{acknowledgments}

\section*{Data Availability}
The data that support the findings of this study are available from the corresponding author upon reasonable request.
\bibliographystyle{unsrtnat}
\bibliography{nemd_ref}
\end{document}